\documentclass[journal]{IEEEtran}
\usepackage{mathtools}
\ifCLASSINFOpdf
\usepackage{subfig}
\usepackage[pdftex]{graphicx}
\else
\fi
\usepackage{algorithmic}
\begin{document}
\title{
Evolutionary Stability of Reputation Management System in Peer to Peer Networks
}
\author{Antriksh Goswami,
        Ruchir Gupta% <-this % stops a space
%\thanks{Antriksh Goswami is the PhD student with the Department
%of Computer Science and Engineering, Indian Institute of Information Technology, Jabalpur, 482005 India e-mail: (g.antriksh@iiitdmj.ac.in).}% <-this % stops a space
%\thanks{Ruchir Gupta are with Anonymous University.}% <-this % stops a space
%\thanks{Manuscript received April 19, 2005; revised August 26, 2015.}
}
\maketitle
\begin{abstract}
Each participant in peer-to-peer network prefers to ”free-ride” on the contribution of other participants. Reputation based resource sharing is a way to control the free riding. Instead of classical game theory we use evolutionary game theory to analyse the reputation based resource sharing in peer to peer system. Classical game-theoretical approach requires global information of the population. However, the evolutionary games only assumes light cognitive capabilities of users, that is, each user imitates the behavior of other user with better payoff. We find that without any extra benefit reputation strategy is not stable in the system. We also find the fraction of users who calculate the reputation for controlling the free riding in equilibrium. In this work first we made a game theoretical model for the reputation system and then we calculate the threshold of the fraction of users with which the reputation strategy is sustainable in the system. We found that in simplistic conditions reputation calculation is not evolutionarily stable strategy but if we impose some initial payment to all users and then distribute that payment among the users who are calculating reputation then reputation is evolutionary stable strategy.
\end{abstract}
\IEEEpeerreviewmaketitle
\section{Introduction}
\IEEEPARstart{P}{eer-to-peer} systems are autonomous and distributed dynamic resource-sharing networks. Collectively, the resources of many autonomous users builds an economic and highly scalable platform for data-sharing, storage and distributed computing etc. In these systems, it is peremptory for peers to voluntarily contribute resources which includes storage, bandwidth and data content etc. However, instinctively, each peer would prefer to ''free ride'' on the part of other peers by consuming available resources and services without contributing anything back, and thus avoid the corresponding costs. It was reported that nearly 70\% of Gnutella users share nothing with other users (these users simply free-ride on other users who share information),
and nearly 50\% of all file search responses come from the top 1\% of information sharing nodes \cite{adar2000free}. In a follow-up study (5 yr
later), it was found that 85\% of users share nothing \cite{hughes2005free}, which implies the free-riding problem had got worse in the intervening years.

 Generally, the lack of cooperation and so free riding is a major problem in these autonomous resource sharing networks \cite{karakaya2009free}, \cite{karakaya2008counteracting}, \cite{feldman2005overcoming}. Designer of P2P system can consider either of two ways for resource management in these systems: resource allocation in which the designer should decide whether and what percentage of a good (with given predefined capacity) each peer should consume and resource provision in which the designer's task is to entice independent participant to provide resource (with its right share). 
 
 Both mechanisms in a way or other use the reputation of the peers. Ideally, reputation should be the measure of cooperative behavior of a node which is an abstract quantity and a private information of a node. So, it is difficult to measure the cooperative behavior of a node and we
can only measure its implications with some degree of uncertainty. However, it can be estimated with certain
accuracy on the basis of behavior observed by a node. A number of mechanisms have been proposed
in literature \cite{buragohain2003game, lee2003cooperative, dutta2003design, papaioannou2006reputation, andrade2004discouraging, marti2004limited, kamvar2003eigentrust, xiong2004peertrust, zhou2007powertrust, zhou2008gossiptrust} for calculation of
reputation.

A considerable amount of work has already been done on resource allocation \cite{satsiou2010reputation}, \cite{gupta2015reputation}  and resource provisioning \cite{wang2015vpef} using reputation. Reputation based resource allocation mechanisms play a
crucial role to encourage cooperation among autonomous nodes. Research till now reveals that reputation calculation is useful to entice the cooperation, but the evolutionary stability of the reputation system is yet to be investigated.

In this work, we have analyzed the reputation system for understanding the conditions of evolutionary stability of the system. Reputation systems always tend to give more benefit to those nodes which are contributing more to the system. But as calculation of reputation of the node requires some cost so, cooperators ($C$ strategy users) who are not calculating reputation always gets more benefit when they interact with reputation calculators ($R$) as compare to $R$ users when they interact with another $R$ users. Due to this $R$ strategy is not evolutionary stable strategy if there is not extra benefit. In this work, we have analyzed a payment based mechanism which gives the required benefit to the reputation strategy so that it could be evolutionary stable. 
%We in this work consider three types of strategic nodes, the first one is Reputation Calculation with cooperation (R), this strategic node always provide requested services as per the reputation of the requesters and also calculate the reputation of the user, second is Cooperators (C), who always provide the requested services to all users, third is Defection (D), who always deny any requested service. R strategic nodes always pay some cost for calculation of reputation, due to this cooperators (C strategy users) who are not calculating reputation always gets more benefit when they interact with reputation calculators (R) as compare to R users when they interact with another R users. Due to this R strategy is not evolutionary stable strategy if there is not extra benefit. In this work we propose a payment based mechanism which gives the required benefit to the reputation strategy so that it could be evolutionary stable.  
In our work, for the sake of simplicity we considered only discrete value full contribution, full defection and reputation calculation with full contribution as a strategic choice. 
Main findings of this paper are listed as follows: 
\begin{enumerate}
\item The threshold of the number of `reputation calculator' $R$ strategy users which if keep fixed then cooperators always gets higher payoff than defectors and so `free riding' can be controlled.
\item The threshold of the number of $R$ strategy users which if keep fixed then reputation strategy users always gets higher payoff than defectors.
\item If we allow reputation as the optional strategy then in general conditions (without any extra benefit to $R$ and $C$ strategy users) $R$ strategy is not an equilibrium strategy.
\item If we impose some initial payment and distribute that initial payment among the players who are calculating the reputation then reputation is evolutionary stable strategy for a threshold of initial payment.
\end{enumerate}
\section{Related Work}
In literature a lot of study has been done for estimation of reputation. Buragohain et.al. \cite{buragohain2003game} take the ratio of resource contributed
by the node to the ratio of absolute measure of contribution, whether they does not discuss the mechanism to measure of contributions of a node by the receiving node. In \cite{lee2003cooperative} the
receiving node computes the trust value of a node on
the basis of received data in the transactions with the
sending node. Duttay et.al.
\cite{dutta2003design} suggest that each node
should provide rating to the other node on the basis of service provided by the user and then this rating is supervised by a group of users. This scheme uses the reputation in the form of rating. In \cite{papaioannou2006reputation} each node calculates reputation of other node on the basis of
service received from the other nodes depending upon
number of transactions done with those nodes, delay in
the transactions and the download speed. Andrade et.al.
\cite{andrade2004discouraging} calculates the reputation of a node by taking the
difference of resources received from and provided to
the node. In \cite{marti2004limited},\cite{kamvar2003eigentrust},\cite{xiong2004peertrust},\cite{zhou2007powertrust} and \cite{zhou2008gossiptrust}, a node adjusts
the reputation of other node on the basis of quality of
transactions with that node. Eigen-Trust \cite{kamvar2003eigentrust} uses sum of
positive and negative ratings, Peer-Trust \cite{xiong2004peertrust} normalises
the rating on each transaction whereas Power-Trust \cite{zhou2007powertrust}
uses Bayesian approach to calculate reputation locally.

Some resource allocation and resource provision schemes using generosity level of the peers has been investigated. Feldman et al.\cite{feldman2004robust} estimated generosity of node as the ratio of the service provided by the node to the service received by the node. Nodes will be served as per their estimated generosity. Kung et al. \cite{kung2003differentiated} proposed selection of a peer for allocation of resource according to its contribution to the network and usage of resources. Nodes desirous to receive resources have to contribute above a certain level to the network. Meo et al. \cite{meo2005rational} model the resource allocation problem as competition among all requesters on the basis of resource request amount. Resource is allocated to the requesters who are demanding least. In this work author asumed all the requesters as generous means: it does not want not to share,
but to share as little as possible.
  Later the term generosity level of the peer is replaced by the reputation of the peer and some new resource allocation and resource provision schemes worked on this. In \cite{satsiou2010reputation} Satsiou et al. proposes the distributed reputation-based system. They propose the algorithm which maximizes requesters satisfactions as well as maximizes the download capacity of the user so as to its utility. In \cite{gupta2015reputation} Gupta et al. uses the probabilistic approach to allocate the resources on the basis of reputation. They argue that by using this scheme nodes
that don't have very good reputation about each other, may
also serve each other at least some amount of resource
with finite probability. For avoiding whitewashing in unstructured peer to peer to network Gupta et al in \cite{gupta2013avoiding} proposes a reputation based resource allocation mechanism in which the initial reputation is adjusted according to the level of
whitewashing in the network. In \cite{lai2003incentives} Lai et al. uses the decision function that takes shared and subjective history of the previous interactions in deciding whether to cooperate or defect with the requester. Ma et al. in \cite{ma2006incentive} proposes a water filling squared bucket algorithm. In which the width of the bucket is the contribution level of the user and the height is the required demand of the user. The allocation is given on the basis of shorter height first. This mechanism ensures the maximization of individual and social utility. In \cite{ma2006demand} Ma et al. allocate the resources to the users on the basis of their contribution level and requested bandwidth. Yan et al. in \cite{yan2007ranking} uses the contribution level as the ranking of the user and allocate the resources on the basis of the ranking of the user.
 
  All of these schemes considered reputation calculation as compulsory for all the nodes and on the basis of reputation they impose their allocation scheme. But we in this work analyze reputation calculation as a strategy of the user and found that whether a lot of resource allocation mechanism has been given but even reputation calculation is not an evolutionary stable strategy. For making reputation as an evolutionary stable strategy we devise a mechanism for the autonomous peer-to-peer system so that it could be an evolutionary stable strategy.
 In \cite{wang2015vpef} VPEF propose Evolutionary Game Theory based mechanism, VPEF (Voluntary Principle and round-based
Entry Fee), to enforce cooperation in the society. In VPEF author modeled the interaction among the users as public goods game, whereas we modeled the interaction as two player strategic game because all the reciprocation in peer-to-peer network are pairwise interaction. Same as in VPEF we also incorporate round based entry fees. VPEF highlights the role of selection of different strategies whereas we highlight the role of stability of strategies against mutation.
In \cite{seredynski2009evolutionary} author, evolutionary game theoretically analysed the reputation strategy in a mobile ad hoc network using simulation and as of their strategic game reputation strategy is not evolutionarily stable strategy. They are minimizing the possibilities of invading of reputation strategy by always defect strategy. As of the anonymous, autonomous and dynamic nature of the peer-to-peer network author in \cite{wang2011effectiveness} proposes the mechanism in which some cooperators first behave like generous,and then like harsh according to peers' current behaviors. One of the weak-point of the above scheme is whether the punisher will dominate the system, but neither punishment nor cooperation is evolutionary stable strategy.
\section{Modeling of Reputation system as a game}
In this paper, we have used peer, user, node and agent interchangeably. Peer-to-peer network has been assumed as pure i.e., without any central server with total $N$ number of peers . We also assume that any two peers in the network can interact with each other.
A P2P system without any punishment and reward mechanism can simply be modeled as famous Prisoners' Dilemma game in which defection always strictly dominates cooperation strategy. The cooperation strategy can only survive in the system when it can dominate the defection strategy. In the reputation based resource allocation mechanisms, reputation is calculated by the peers. Resources are allocated to the resource requesting peers based on their reputation. 

Reputation management is a tool to punish the defectors but on the cost of reputation calculation. Peers prefer to save this additional cost involved in reputation calculation. Therefore, most of the cooperators does not calculate reputation and only cooperate. If the fraction of reputation calculators in the population comes to lower than a threshold, this leads to the domination of defectors and consequent collapse of system. The threshold can be calculated by modeling whole situation as a strategic game. Although, a user can interact with multiple other users at a time but as each interaction is independent from other interactions, hence we can model all these interactions as pairwise interaction game between two users. We model this phenomena as a symmetric simultaneous game where both players make their moves simultaneously.
\subsection{General Reputation Game}
Here peer's strategy may be classified into three types viz. Reputation Calculation with cooperation ($R$), Cooperation ($C$), Defection ($D$). Users playing $R$ strategy always provide requested services as per the reputation of the requesters; Users playing $C$ strategy always provide the requested services to all users; Users playing $D$ strategy always deny any requested service.
Therefore, reputation system is modeled as the strategic game.\\
\textbf{Players}:- User1, User2\\
\textbf{Strategies}:- Reputation Calculation with cooperation ($R$), Cooperation ($C$), Defection ($D$)\\
\textbf{Preferences}
\begin{eqnarray}
\nonumber &&  U_i(A_i,A_{-i})= (^C l_{-i} \cdot (1- ^Rl_{-i}) + ^Cl_{i} \cdot ^Rl_{-i} \cdot ^Cl_{-i}) \cdot d \\\nonumber &&  -  (^Cl_{i} \cdot (1- ^Rl_i) + ^Cl_{-i} \cdot ^Rl_i)\cdot a - ^Rl_i \cdot \alpha + \\ && (^Cl_i \cdot ^Rl_{-i}) \cdot \beta
\end{eqnarray}
where $A_i$ and $A_{-i}$ are the actions of player $i$ and other than player $i$ respectively. $^Cl_i$ is the cooperation level of player $i$ and $^Rl_i$ is the reputation calculation level of player $i$ respectively.\\
For $C$ (cooperation) strategy : $^Cl = 1$ and $^Rl = 0$. Because these users are always cooperating and not calculating reputation. Similarly for R (reputation calculation with cooperation) strategy : $^Cl = 1$ and $^Rl = 1$, for $D$ (defection) strategy : $^Cl = 0$ and $^Rl = 0$\\
In the preference function the first term represents the `benefit of sharing', the benefit of sharing resources can only be obtained by first user when the second user is either cooperator ($C$) or when the first player is either cooperator or reputation calculator user ($C$ and $R$) and second player is reputation calculator ($R$) user. The second term represents the `cost of sharing', the cost of sharing will only be imposed when the player is either cooperator or he is reputation calculator and second player is cooperator. Third term represents the `cost of reputation calculation' which is always incurred when the first user is reputation calculator ($R$) user. Fourth term is the `benefit of reputation increment'. 
  The payoff matrix of the game is illustrated in table \ref{First_Game}. In this matrix, row corresponds to the possible actions of peer A whereas, column corresponds to the possible actions of peer B and the values in each box are the players' payoffs to the action profile to which the box corresponds, with A's payoff listed first. Each first value $a_{ij}$ of this table symbolizes the payoff of A with strategy $S_i$, when B opts for strategy $S_j$. Take the first value $a_{12}$ for instance, the value $d-a-\alpha$
is the payoff of A with R strategy when B opts $C$ strategy where $a$ and $\alpha$ is the cost incurred due to providing the service to the other player and the cost incurred due to calculation of reputation respectively. In this $\alpha<a$ as the `cost of reputation calculation' is always less than `cost of sharing', otherwise $R$ strategy users loss is more than $C$ strategy users when they play with $D$ strategy users and so will always prefer only to cooperate without calculation of reputation.

If a user with $R$ strategy meets a user with $C$ strategy, it will always grant a service to $C$ strategy user and get a service from the $C$ strategy user. Thus, it would obtain a benefit $d-a$. However, to
calculate the reputation of the peers, the user with $R$ strategy has to communicate to the other peers for
information. So it has to bear an extra cost $\alpha$. Therefore, the total payoff
of user with $R$ strategy in this transaction is $d-a-\alpha$. In this $d>a$ as the benefit received by shared data is always greater than the cost of sharing. Now consider the second value $b_{12}$ that is the payoff of A with $C$ strategy when B opts $R$ strategy. If a user with $C$ strategy meets a user with $R$ strategy it will always grant service and get a service from the $R$ strategy user. Thus, it would obtain a benefit $d-a$. However, due to its cooperative behavior its reputation would also increase, so it would get the extra benefit for reputation increment $\beta$. Therefore, the total payoff of a user with $C$ strategy in this transaction is $d-a+\beta$.

\begin{table}[!t]
\renewcommand{\arraystretch}{1.3}
\caption{Simplistic Model}
\label{First_Game}
\centering
\begin{tabular}{c c	c  c}
{ }&{R(B)} & {C (B)} & {D(B)}\\ \hline
{R (A)} & \shortstack{$d-a-\alpha+\beta$,\\ $d-a-\alpha + \beta$} & \shortstack{$d-a-\alpha$,\\$d-a+\beta$} & {$-\alpha$,0}\\ \hline
{C (A)} & {$d-a+\beta$,$d-a-\alpha$} & {$d-a$,$d-a$} & {$-a$,$d$}\\ \hline
{D (A)} & {0,$-\alpha$} & {$d$,$-a$} & { 0,0}\\ \hline
\end{tabular}
\end{table}
\begin{table}[!t]
\renewcommand{\arraystretch}{1.3}
\caption{symbols used in modeling the reputation game}
\centering
\begin{tabular}{c | c}
\hline
{Symbol} & {Definition} \\ \hline
{$d$} & \shortstack{Benefit received by getting the service from the cooperator} \\ \hline
{$a$} & {The cost incurred due to providing the service to the other player}\\ \hline
{$\beta$} & {The benefit received due to improving the reputation}\\ \hline
{$\alpha$} & {The cost incurred due to calculation of reputation}\\ \hline
{$P_R$} & {Average Payoff of $R$ strategic Players}\\ \hline
{$P_C$} & {Average Payoff of C strategic Players}\\ \hline
{$P_D$} & {Average Payoff of D strategic Players}\\ \hline
{$x_i$} & {The proportion of peers with strategy $i$}\\ \hline
{$n_d$} & {Total number of D strategy users}\\ \hline
{$n_r$} & {Total number of $R$ strategy users}\\ \hline
{$n_c$} & {Total number of C strategy users}\\ \hline
{$p$} & { Round based payment payed by users}\\ \hline
\end{tabular}
\end{table}
\textbf{Analysis}:- In this game if $R$ strategy user interacts with $D$ strategy user, then he gets payoff $-\alpha$ and if $C$ strategy user interacts with $D$ strategy user, then he gets payoff $-a$ which are less than $0$. This shows that users does not have higher payoff in unilaterally deviation from profile $(D,D)$. Therefore $(D,D)$ is the pure strategic strict Nash equilibrium and consequently $D$ is the evolutionary stable strategy (ESS).
\begin{equation}
U_{1}(D,D) > U_{1}(x,D)
\end{equation}
where $x$ is any strategy other than $D$. \\
Let us assume that the population fraction of $R$, $C$ and $D$ strategies are $x_{R}$, $x_{C}$, and ($1-x_{R}-x_{C}$) respectively. Therefore, the average payoff of each strategy is written as
\begin{eqnarray}
P_{R} &=& d \cdot (x_{R}+x_{C}) - a \cdot(x_{R}+x_{C}) + x_{R} \cdot \beta - \alpha\\
P_{C} &=& d \cdot (x_{R}+x_{C}) - a + x_{R} \cdot \beta\\
P_{D} &=& x_{C} \cdot d.
\end{eqnarray}
From the above equations following observations can be made.
\begin{itemize}
\item If the reputation calculation cost $\alpha$ is assumed to be negligible, then the expected payoff for $R$ strategy users will always be greater than the $C$ strategy users till there are $D$ strategy users as in this case $x_{R}+x_{C}$ is less than 1 and it will be equal when there is no $D$ strategy user
\item The payoff received by $R$ strategy will be higher than the $D$ strategy i.e., $P_{R}> P_{D}$ when $x_{R} > \frac{a\cdot(x_{R}+x_{C}) +\alpha } {(d+\beta)}$ i.e., when fraction of $R$ strategy user is greater than the ratio of total expected cost incurred to $R$ strategy users by population and individual benefit received by $R$ strategy user when played with $R$ strategy user. The payoff received by $C$ strategy will be higher than the $D$ strategy i.e., $P_{C} > P_{D}$ when $x_{R} > \frac{a}{(d+\beta)}$ i.e., when the fraction of $R$ strategy user is greater than the ratio of total expected cost payed by $C$ strategy user and individual benefit received by $C$ strategy user when played with $R$ strategy user.
\item If fraction of $R$ strategy users are lesser than both the ratio mentioned above, then the payoff of $D$ strategy users becomes highest in the population and therefore users imitates to $D$ strategy, because now $P_{D}>P_{R}$ and $P_{D} >P_{C}$.
\item The payoff to $R$ strategy users will be higher than $C$ and $D$ strategy when $P_{R}>P_{C}$ and $P_{R}>P_{D}$ i.e., $x_{D} > \frac{\alpha}{a}$ i.e., when the fraction of $D$ strategy users is greater than the ratio of cost of reputation calculation and cost of sharing, and also when $x_{R} > \frac{a\cdot(x_{R}+x_{C}) +\alpha } {(d+\beta)}$.
\end{itemize}
We have already examined pure strategy equilibrium now let us examine the mixed strategy equilibriums of the game.

\textbf{Existence of mixed strategy Nash Equilibrium} For the mixed strategy equilibrium first we will examine the mixed strategy with any two strategies, then we will take the combination of all three strategies. 

Let us consider the combination of two strategies. First take $C$ and $R$ strategy. In this combination $C$ always dominates the $R$ strategy which is then dominated by the $D$ strategy. If we take $C$ and $D$ strategy, then $D$ always dominates the $C$ strategy. If we take $R$ and $D$ strategy, then although in this combination $R$ and $D$ strategy is in itself pure strategy Nash equilibrium but only $D$ strategy fulfills the condition for pure strategy Nash equilibrium in the presence of $C$ strategy. This two strategy combination provides another mixed strategy Nash equilibrium with zero payoff. The equilibrium can be obtained by solving following equations.
\begin{eqnarray}
x_{R}\cdot (d-a-\alpha+\beta) + (-\alpha) \cdot (1-x_R)=0  
\end{eqnarray}
By this equality we got
\begin{subequations}
\label{MNE2_game1}
\begin{align}
x_R &= \frac{\alpha}{(d-a+\beta)}\\
x_C &= 0\\
x_D &= 1-\frac{\alpha}{(d-a+\beta)}
\end{align}
\end{subequations}
The second mixed strategy equilibrium \ref{MNE2_game1} is only possible when the payoff of $C$ strategy with the above combination is less than or equal to the payoff of $R$ and $D$ strategy users i.e., $P_C <= 0 $ i.e., $(d-\beta) \geq \frac{a^2}{a-\alpha}$.
This equilibrium leads to zero payoff so this is not useful from system designer perspective. Now we will analyze the mixed strategy equilibrium with all three strategy.

 For mixed strategic equilibrium with all three strategies:\\
\begin{eqnarray}
\label{game1_first}
\nonumber && x_{R} \cdot (d-a-\alpha+\beta) + x_{C} \cdot (d-a-\alpha) + (1-x_{R}-x_{C}) \cdot\\\nonumber && (- \alpha)
 = x_{R} \cdot (d-a+\beta) + x_{C} \cdot (d-a) + (1-x_{R}-x_{C}) \cdot\\ && (-a)
= x_{R} \cdot 0 + x_{C} \cdot d + (1-x_{R}-x_{C}) \cdot 0
\end{eqnarray}
By this equality we got 
\begin{subequations}
\label{MNE1_game1}
\begin{align}
 x_{R} &= \frac{a}{d+\beta} \\
 x_{C} &= \frac{(d+\beta)(a-\alpha)- a^2}{(d+\beta) \cdot a}\\
 x_{D} = 1-x_{R}-x_{C} &= \frac{\alpha}{ a}
  \end{align}
\end{subequations}
\subsubsection{Theoretical Analysis of Mixed Strategy Equilibrium with All Three Strategies}
In this game, ($D$,$D$) is a strict Nash equilibrium but this equilibrium state leads to no sharing from all the peers and results in collapse of the system. The only equilibrium state which allow the survival of the system is polymorphic mixed strategy equilibrium depicted by equation (\ref{MNE1_game1}).  In this section we analyzed the mixed strategy equilibrium equation for varying a single parameter value (viz. $d$, $a$, $\alpha$, $\beta$) while other parameters remain fixed.\\ By this analysis of the mixed strategy equilibrium in equation \ref{MNE1_game1} following things can be observed:-
\begin{itemize}
\item With the increment in `cost of the reputation calculation' $\alpha$, $D$ strategy users increases, $C$ strategy users decreases and $R$ strategy users remains same in resulting mixed strategy equilibrium
\item With the increment in `cost of sharing' $a$ , $R$ strategy users increases, $D$ strategy users decreases.
\item With the increment in `benefit of sharing' $d$, $C$ strategy users increases, $R$ strategy users decreases and $D$ strategy users remains same in resulting mixed strategy equilibrium.
\item With the increment in `benefit of reputation increment' $\beta$ the fraction of $R$ strategy users decreases, fraction of $C$ strategy users decreases and fraction of $D$ strategy users remains constant.
\end{itemize}
\begin{figure}
  \subfloat[]{%
  \begin{minipage}{0.5\linewidth}
  \label{dVsF1G}
  \includegraphics[width=1\linewidth]{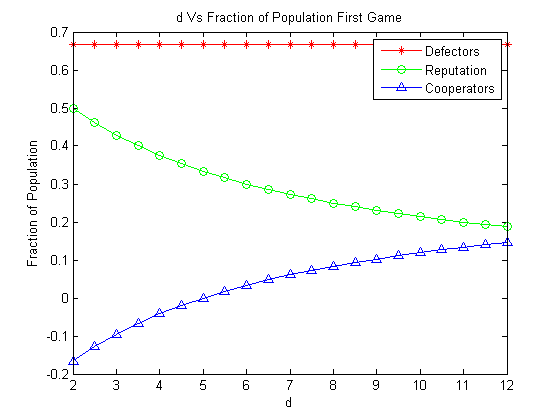}\hfill
  \end{minipage}%
  }
  \subfloat[]{%
  \begin{minipage}{0.5\linewidth}
  \label{alphaVF1G}
  \includegraphics[width=1\linewidth]{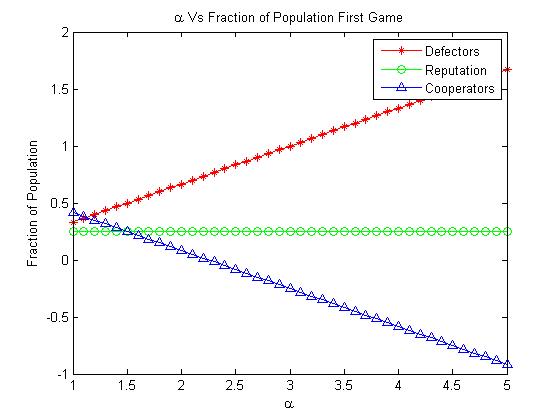}\hfill
  \end{minipage}%
  }\par
  \subfloat[]{%
  \begin{minipage}{0.5\linewidth}
  \label{aVF1G}
  \includegraphics[width=1\linewidth]{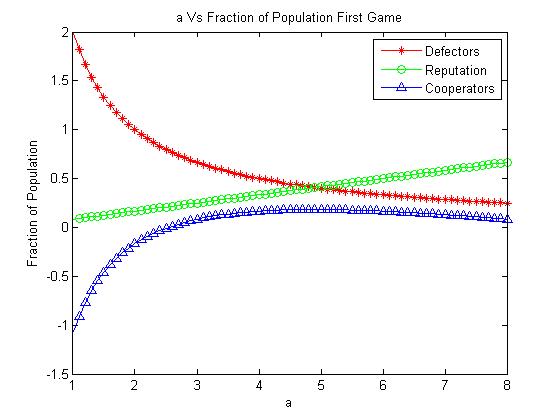}
  \end{minipage}%
  }
  \subfloat[]{%
    \begin{minipage}{0.5\linewidth}
    \label{betaVF1G}
    \includegraphics[width=1\linewidth]{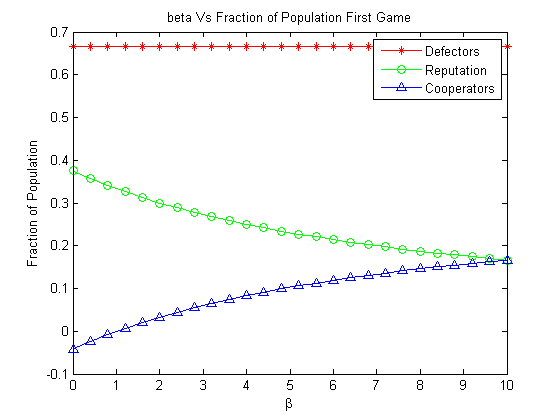}\hfill
    \end{minipage}%
    }\par
  \caption{Fraction of population in mixed Nash equilibrium for the network (a) vs benefit of shared data $d$ with other parameters\{ $a=3$, $\alpha=3$, $\beta=4$, $p=0.5$\} (b)vs cost of reputation calculation $\alpha$ with other parameters \{$d=8$, $a=3$, $\beta=4$, $p=0.5$\} (c)vs cost of sharing $a$ with other parameters \{$d=8$, $\alpha=2$, $\beta=4$, $p=0.5$\}}
\end{figure}
The reasoning behind first observation is that as `cost of reputation calculation' $\alpha$ increases, then the payoff to $R$ strategy users decreases and therefore the $R$ strategy becomes less lucrative to the users so they switches to $C$ strategy and $D$ strategy users. As $R$ strategy users decreases and $C$ strategy users increases, then the payoff to $C$ strategy users also decreases and payoff to $D$ strategy users increases. Due to this the $C$ strategy users also switches to $D$ strategy users. As $D$ strategy users increases the payoff to $C$ strategy users decreases more and comes to lower than $R$ strategy. Due to this $C$ strategy users now switches to $R$ strategy till the payoff of all the strategy equalizes. Due to this in new equilibrium $D$ strategy users increases, $C$ strategy users decreases and $R$ strategy users remains same as shown in figure \ref{alphaVF1G}.
The reasoning behind second observation is that as `cost of sharing' $a$ increases, then the payoff of $R$ and $C$ strategy users decreases. Due to this most of the $R$ and $C$ strategy users switches to $D$ strategy users. As $C$ strategy users decreases the payoff to $D$ strategy also decreases and so $D$ strategy users also switches to $R$ strategy. As $R$ strategy users increases the payoff to $C$ strategy users increases and so $D$ strategy users also switches to $C$ strategy. Due to this in final equilibrium $R$ strategy users increases, $D$ strategy users decreases.

  The reasoning behind third observation is, at one equilibrium state when the expected payoff of all the three strategies are same, then as benefit of shared data increases the expected payoff of $R$ user and $C$ user increases equally whereas the expected payoff of $D$ user increases $x_{R}$ fraction lesser than $R$ and $C$ strategy users. Due to this $D$ strategy users switches to $C$ strategy users and $R$ strategy users equally. But as the fraction $x_{R}$ and $x_{C}$ increases, then the expected payoff of reputation $R$ users increases lesser than expected payoff of $C$ strategy users and as fraction of $C$ strategy users increases the expected payoff to $D$ strategy users also increases. So $R$ strategy users switches to cooperation and defection users. As $x_{R}$ decreases the payoff to $C$ strategy users also decreases so $C$ strategy users switches to $D$ strategy. This drift process comes with next equilibrium state in which $R$ strategy users decreases, $C$ strategy users increases and $D$ strategy users remains same as shown in figure \ref{dVsF1G}.
  
  The reasoning behind fourth observation is same as the `benefit of shared resources'. In this as the `benefit of reputation increment' $\beta$ would increase the expected payoff of reputation and $C$ strategy users would increase whereas the expected payoff of $D$ strategy users remain same. Due to this $D$ strategy users switches to $R$ and $C$ strategy users. As the $C$ strategy users increases the payoff of $D$ strategy users again increases. As the fraction of reputation and cooperation increases the expected payoff of $R$ users increases lesser than $C$ strategy. So $R$ strategy users switches to $D$ and $C$ strategy users and in next equilibrium $D$ strategy users remain unchanged and fraction of $R$ strategy users decreases and fraction of $C$ strategy users increases.
\subsection{Reputation system with round based initial payment distributed to cooperative users}
In the previous subsection, we have analyzed the reputation system in which there is no initial payment required for the peers. Now in this subsection we will analyze the reputation system with round based initial payment imposed to the peers, which is distributed among the cooperation and reputation with cooperation peers ($C$ and $R$ strategy). The game can be defined as follows.\\
\textbf{Players}:- User1, User2\\
\textbf{Strategies}:- Reputation with cooperation ($R$), Cooperation ($C$), Defection (D)\\
\textbf{Preferences}
\begin{eqnarray}
\nonumber &&  U_i(A_i,A_{-i})= (^C l_{-i} \cdot (1- ^Rl_{-i}) + ^Cl_{i} \cdot ^Rl_{-i} \cdot ^Cl_{-i}) \cdot d \\\nonumber && -  (^Cl_{i} \cdot (1- ^Rl_i) + ^Cl_{-i} \cdot ^Rl_i)\cdot a - ^Rl_i \cdot \alpha + \\\nonumber && (^Cl_i \cdot ^Rl_{-i}) \cdot \beta + ^Cl_i \cdot \frac{N \cdot p}{N-n_d} -p
\end{eqnarray}
where $A_i$ and $A_{-i}$ are the actions of player $i$ and player other than $i$ respectively. $^Cl_i$ is the cooperation level of player $i$ and $^Rl_i$ is the reputation calculation level of player $i$ respectively.\\
For $C$ (cooperation) strategy : $^Cl = 1$ and $^Rl = 0$. Because these users are always cooperating and not calculating reputation. Similarly for $R$ (reputation calculation) strategy : $^Cl = 1$ and $^Rl = 1$ and for $D$ (defection) strategy : $^Cl = 0$ and $^Rl = 0$\\
In the preference function the first term represents the benefit of sharing, the benefit of sharing the resources can only be obtained by first user when the second user is either cooperator ($C$) or when the first player is either cooperator or reputation calculator user ($C$ and $R$) and second player is reputation calculator ($R$) user. The second term represents the `cost of sharing', the cost of sharing will only be imposed when the player is either cooperator or he is reputation calculator and second player is cooperator. Third term represents the cost of reputation calculation cost which is always incurred when the first user is reputation calculator ($R$) user. Fourth term is the benefit of reputation increment. Fifth term is the benefit due to initial payment distribution. Payoff matrix of the game is shown in table \ref{Second_Game}.
\begin{table}[!t]
\renewcommand{\arraystretch}{1.3}
\caption{Initial Payment With Distribution To Cooperative Users}
\label{Second_Game}
\centering
\begin{center}
\begin{tabular}{c c	c  c}
{ }&{R(B)} & {C (B)} & {D(B)}\\ \hline
{R (A)} & \shortstack{$d-a-\alpha+\beta+$\\ $\frac{n_d \cdot p}{n_r+n_c}$, \\$d-a-\alpha + \beta +$ \\ $\frac{n_d \cdot p}{n_r+n_c}$} & \shortstack{$d-a-\alpha +$ \\$ \frac{n_d \cdot p}{n_r+n_c}$,\\$d-a+\beta+$ \\ $\frac{n_d \cdot p}{n_r+n_c}$} & {$\frac{n_d \cdot p}{n_r+n_c} - \alpha$,$-p$}\\ \hline
{C (A)} & \shortstack{$d-a+\beta+$\\ $\frac{n_d \cdot p}{n_r+n_c}$,\\$d-a-\alpha+$\\ $\frac{n_d \cdot p}{n_r+n_c}$} & \shortstack{$d-a+\frac{n_d \cdot p}{n_r+n_c}$,\\$d-a+\frac{n_d \cdot p}{n_r+n_c}$} & \shortstack{$-a+\frac{n_d \cdot p}{n_r+n_c}$,\\$d-p$}\\ \hline
{D (A)} & {$-p$,$\frac{n_d \cdot p}{n_r+n_c} - \alpha$} & \shortstack{$d-p$,\\$-a+\frac{n_d \cdot p}{n_r+n_c}$} & { $-p$,$-p$}\\ \hline
\end{tabular}
\end{center}
\end{table}

\textbf{Analysis}
In this game we claim that
\begin{equation}
\label{G2NSScond}
\begin{split}
if\ \ \ \ p \geq a \cdot (1-\frac{n_d}{N})
\end{split}
\end{equation}\\
then cooperation strategy profile i.e., ($C$,$C$) will be the Nash equilibrium and
\begin{equation}
\label{G2ESScond}
\begin{split}
if\ \ \ \ p > a \cdot (1-\frac{n_d}{N})
\end{split}
\end{equation}\\
then cooperation strategy profile will be Evolutionarily stable strategy.\\
If the condition in equation \ref{G2NSScond} is not fulfilled and
\begin{equation}
\label{G2ND}
\begin{split}
if\ \ \ \ p < \alpha \cdot (1-\frac{n_d}{N})
\end{split}
\end{equation}\\
then the defection strategy profile i.e., ($D$,$D$) will be the pure strategy Nash equilibrium profile.

The argument for strategy profile ($C$,$C$) as Nash equilibrium, given condition (\ref{G2NSScond}), is as follows.\\
We can observe that if condition (\ref{G2NSScond}) is true, then $U_1(C,C) \geq U_1(D,C)$ and $U_1(C,C) \geq U_1(R,C)$ that means the payoff of cooperation strategy when played with itself is always greater than or equal to other two strategy while played with cooperation.\\
The argument for strategy profile ($C$,$C$) as Evolutionary Stable, given condition (\ref{G2ESScond}), is as follows.\\
We can observe that if condition (\ref{G2ESScond}) is true, then $U_1(C,C) > U_1(D,C)$ and $U_1(C,C) > U_1(R,C)$ that means the payoff of cooperation strategy when played with itself is always strictly greater than $D$ and $R$ strategy while played with $C$ strategy.\\
The argument for strategy profile ($D$,$D$) as Nash equilibrium as well as Evolutionary stable, given condition (\ref{G2ND}), is as follows\\
We can observe that if condition (\ref{G2ND}) is true, then $U_1(D,D) > U_1(R,D)$ and $U_1(D,D) > U_1(C,D)$ that means the payoff of Defection strategy when played with itself is always greater than R and $C$ strategy when played with $D$ strategy.

If condition \ref{G2NSScond} and \ref{G2ND} is not satisfied, then there is no pure strategic Nash equilibrium in this game. Therefore, now we will compute the mixed strategy Nash equilibrium profile. For this the expected payoff of each strategy can be written as\\
\begin{subequations}
\begin{align}
\label{PayoffR}
P_{R} &=d \cdot (x_{R}+x_{C}) - a \cdot(x_{R}+x_{C}) + x_{R} \cdot \beta - \alpha + \frac{n_d \cdot p}{N-n_d}\\
\label{PayoffC}
P_{C} &=d \cdot (x_{R}+x_{C}) + x_{R} \cdot \beta - a  + \frac{n_d \cdot p}{N-n_d}\\
\label{PayoffD}
P_{D} &=x_{C} \cdot d - p
\end{align}
\end{subequations}

 First we will consider the mixed strategy in the combination of two strategies. If we take the combination of only $R$ and $C$ strategies, then in this combination $C$ strategy always dominates $R$, so no mixed strategy equilibrium exist. Now we consider the combination of $R$ and $D$ strategies, then if condition (\ref{G2ND}) is not fulfilled, this results in the domination of $R$ strategy over $D$ strategy and hence again mixed strategy equilibrium does not exist. But if condition (\ref{G2ND}) is fulfilled, then there exist a mixed strategy Nash equilibrium which can be obtained by equating the expected payoffs of $R$ and $D$ strategy users such that,
\begin{eqnarray}
\nonumber && x_{R} \cdot (d-\alpha - a + \beta + \frac{n_d \cdot p}{N-n_d}) + (1-x_R)\cdot\\\nonumber &&(\frac{n_d \cdot p}{N-n_d} - \alpha) = -p\\ && x_R \cdot (d-a+\beta) = \alpha - \frac{p \cdot N}{N-n_d}
\end{eqnarray}
\begin{subequations}
\begin{align}
x_R & = \frac{\alpha - p\cdot(\frac{N}{N-n_d})}{(d-a+\beta)}\\
x_C & = 0\\
x_D & = 1-\frac{\alpha - p\cdot(\frac{N}{N-n_d})}{(d-a+\beta)}
\end{align}
\end{subequations}
This equilibrium leads to negative payoff so this is not useful from system designer perspective.

Now we consider the combination of $C$ and $D$ strategy. If condition \ref{G2ESScond} is fulfilled, then $C$ strategy dominates $D$ strategy and so no mixed strategy equilibrium presents. But if $p = a \cdot (1-\frac{n_d}{N})$, then $U_1(C,C) = U_1(D,C)$ and $U_1(C,D) = U_1(D,D)$. So at this condition although cooperation is pure strategic weak Nash equilibrium, but as $D$ strategy users are also getting the same payoff so there exist a mixed strategy Nash equilibrium which can be obtained by equating the payoff of $D$ and $C$ strategy users such that,
\begin{eqnarray}
\nonumber && x_C \cdot (d-a+\frac{n_d \cdot p}{N-n_d}) + (1-x_C)\cdot(-a+\frac{n_d \cdot p}{N-n_d}) =\\
&& x_C \cdot (d-p) + (1-x_C)\cdot(-p)
\end{eqnarray}
this drift would be there till the payoff of reputation strategy is lesser than these two strategies i.e., 
\begin{equation}
\begin{split}
 x_C \cdot (d-a+\frac{n_d \cdot p}{N-n_d}) + (1-x_C)\cdot(-a+\frac{n_d \cdot p}{N-n_d}) >\\
 x_C \cdot (d-a-\alpha+\frac{n_d \cdot p}{N-n_d})  + (1-x_C) \cdot (-\alpha +\frac{n_d \cdot p}{N-n_d} )
\end{split}
\end{equation}
By solving above inequality we get
\begin{equation}
x_D < \frac{\alpha}{a}
\end{equation}
This shows that till the fraction of defectors remains lesser than the ratio of reputation cost and cost of sharing, reputation users will not be there in the system. This is because when the defectors are less in the society, then paying the reputation cost seems less useful. But as defectors increase, the payoff to reputation strategy increases and users mutates to the reputation strategy. 

Now we will find out the mixed strategy with all the three strategies. For this equilibrium, the expected payoff to all three strategies should be equal.
\begin{eqnarray}
\nonumber && (d-a-\alpha + \beta + \frac{n_d \cdot p}{N-n_d})\cdot x_{R} + (d-a-\alpha \\ 
\nonumber && + \frac{n_d \cdot p}{N-n_d}) \cdot x_{C} + (\frac{n_d \cdot p}{N-n_d} - \alpha)(1-x_{R}-x_{C}) \\
\nonumber && = (d-a + \beta +\frac{n_d \cdot p}{N-n_d})\cdot x_{R} + (d-a+\frac{n_d \cdot p}{N-n_d}) \cdot x_{C} \\ \nonumber && + (-a + \frac{n_d \cdot p}{N-n_d})(1-x_{R}-x_{C}) \\
&& =(-p) \cdot x_{R} + (d-p) \cdot x_{C} + (1-x_{R}-x_{C}) \cdot (-p)
\end{eqnarray}
By solving above equality
\begin{subequations}
\label{MNE3_game2}
\begin{align}
x_{D} &= \frac{\alpha}{a}\\
x_{R} &= \frac{(a-p \cdot (\frac{N}{N-n_d}))}{d+\beta} \\
x_{C} &= \frac{(a-\alpha)}{a} - \frac{a-p \cdot (\frac{N}{N-n_d})}{(d+\beta)}
\end{align}
\end{subequations}
Putting this fraction of $D$ strategy users in the fraction of reputation and cooperation strategy we got
\begin{subequations}
\begin{align}
x_{R} &= \frac{(a-p \cdot (\frac{a}{a-\alpha}))}{d+\beta} \\
x_{C} &= \frac{(a-\alpha)}{a} - \frac{a-p \cdot (\frac{a}{a-\alpha})}{(d+\beta)}
\end{align}
\end{subequations}

\begin{figure}
  \subfloat[]{%
  \begin{minipage}{0.5\linewidth}
  \label{aVF2G}
  \includegraphics[width=1\linewidth]{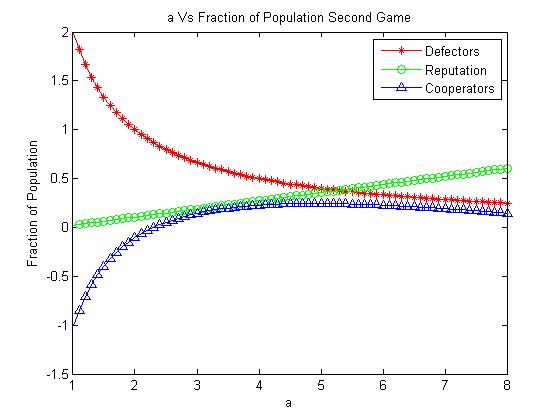}\hfill
  \end{minipage}%
  }
  \subfloat[]{%
  \begin{minipage}{0.5\linewidth}
  \label{alphaVF2G}
  \includegraphics[width=1\linewidth]{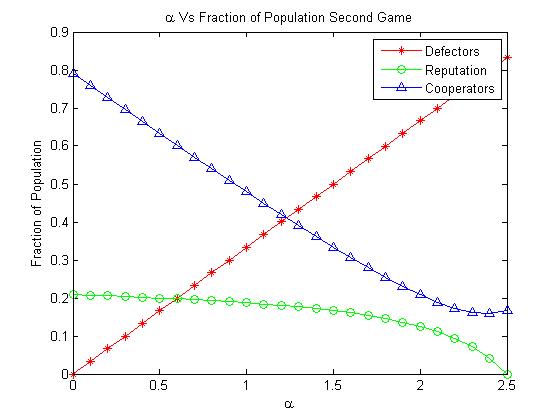}\hfill
  \end{minipage}%
  }\par
  \subfloat[]{%
  \begin{minipage}{0.5\linewidth}
  \label{betaVF2G}
  \includegraphics[width=1\linewidth]{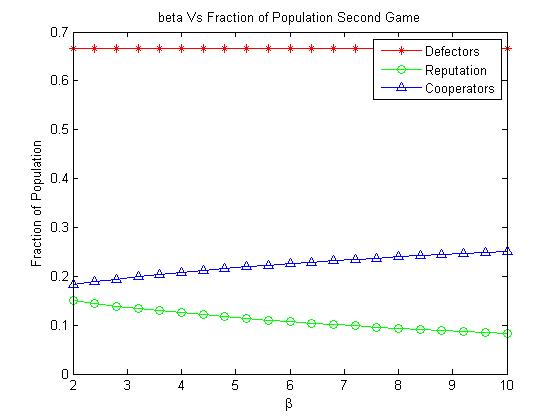}
  \end{minipage}%
  }
   \subfloat[]{%
  \begin{minipage}{0.5\linewidth}
  \label{pVF2G}
  \includegraphics[width=1\linewidth]{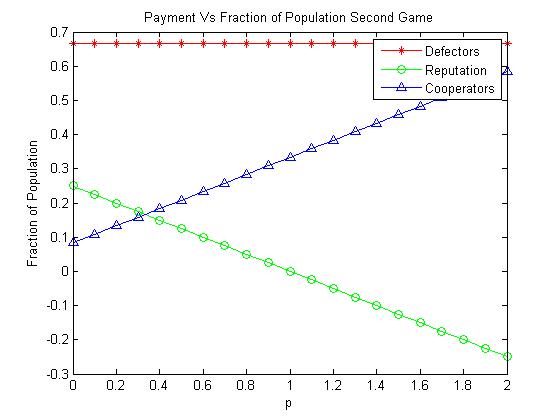}\hfill
  \end{minipage}%
  }\par
  \caption{parameter vs Fraction of population in mixed Nash equilibrium for the network with (a) vs cost of sharing $a$ with other parameters \{$d=8$, $\alpha=2$, $\beta=4$, $p=0.5$\} (b) vs cost of reputation calculation $alpha$ with other parameters \{$d=8$, $a=3$, $\beta=4$, $p=0.5$\} (c) vs benefit of reputation increment with other parameters \{ $d=8$, $\alpha=2$, $a=3$, $p=0.5$\} (d) vs round based initial payment $p$ with other parameters \{$\beta=4$, $\alpha=2$, $a=3$, $d=8$\}}
\end{figure}

In this mixed strategy equilibrium described by (\ref{MNE3_game2}), following things can be observed:-
\begin{itemize}
\item With the increment in `cost of the reputation calculation' $\alpha$, $D$ strategy users increases, $C$ and $R$ strategy users decreases in resulting mixed strategy equilibrium
\item With the increment in `cost of sharing' $a$, $R$ strategy users increases whereas $D$ strategy users decreases in resulting mixed strategy equilibrium
\item The fraction of $R$ strategy is inversely proportional to `initial payment' $p$ whereas the fraction of $C$ strategy users is directly proportional to $p$. Moreover, the mixed strategy equilibrium is not defined for $p$ greater than $a\cdot(1-x_D)$
\item The fraction of $R$ strategy users decreases, fraction of $C$ strategy users increases and fraction of $D$ strategy users remain same with the increment in `benefit of reputation increment' $\beta$
\end{itemize}
The reasoning of first observation is, as the `cost of the reputation calculation' $\alpha$ increases, the expected payoff to $R$ strategy users decreases so they switches to cooperation and defection users. As the cooperation increases the payoff to $D$ strategy users increases and as $R$ strategy users decreases the payoff to $C$ strategy users also decreases so $C$ strategy users also switches to $D$ strategy. As $D$ strategy users increases the payoff to reputation and cooperation users slightly increases because now they are getting benefit of the payment $p$. As defection increases and cooperation decreases the payoff to defection also decreases and so they switches to $R$ strategy users. As $R$ strategy users increases the payoff to $C$ strategy users increases so some $D$ strategy users now switches to cooperation. This whole process shifts the equilibrium where $x_{R}$ and $x_{C}$ decreases and $x_{D}$ increases. Unlike previous game in this game fraction of $R$ strategy users decreases as $\alpha$ increases because as defection increases the payoff to cooperation also increases due to payment so some $R$ strategy users switches to cooperation in equilibrium as shown in figure \ref{alphaVF2G}\\
The reasoning behind second observation is, as `cost of sharing' $a$ increases, the expected payoff to $R$ and $C$ strategy users decreases, but the payoff to $C$ strategy users decreases more so $C$ strategy users switches to $D$ strategy users which currently has highest payoff. As the fraction of $C$ strategy users decreases the expected payoff to $D$ strategy users also decreases and it comes to lowest. So now $D$ strategy users switches to $C$ strategy users and $R$ strategy users. Like previous game in this game the next equilibrium comes at point where fraction of $R$ strategy users increases and fraction of $D$ strategy users decreases but as compare to previous game the rate of change is low because now the $R$ strategy users also getting benefit of the payment from $D$ strategy users.\\

The reasoning behind third observation is, as the `initial payment' $p$ increases, the expected payoff of $C$ and $R$ strategy users increases whereas the expected payoff to $D$ strategy users($d \cdot x_{C} -p$) decreases. As the payoff to $R$ and $C$ strategy users increases the $D$ strategy users switches to $R$ and $C$ strategy users. As $D$ strategy users decreases and $R$ strategy users increases the payoff to $C$ strategy users increases more than $R$ strategy users due to this $R$ strategy users switches to $C$ strategy users. As $R$ strategy users decreases and $C$ strategy users increases the payoff to $D$ strategy users increases so some $R$ and $C$ strategy users switches to $D$ strategy. Due to this process fraction of $R$ strategy users decreases, $C$ strategy users increases and $D$ strategy users remains same in new Nash equilibrium as shown in Figure \ref{pVF2G}. If the initial payment satisfies the condition \ref{G2ESScond} i.e., C strategy is evolutionary stable, then mixed strategy equilibrium is not defined and hence, it can be observed in figure \ref{pVF2G} that the $R$ strategy users fraction becomes negative for initial payment greater than $a\cdot(1-x_D)$ i.e., $0.99$.

The reasoning behind fourth observation is, as the `benefit of reputation' $\beta$ increases the expected payoff of $C$ and $R$ strategy users increase equally whereas the expected payoff of $D$ strategy users remain same. Due to this $D$ strategy users switches to $R$ and $C$ strategy users. As the fraction of $R$ strategy users increases and $D$ strategy users decreases, the payoff of $C$ strategy users increases more than $R$ strategy users so $R$ strategy users also switches to $C$ strategy. As $C$ strategy users increases the payoff of $D$ strategy users again increases and $C$ and $R$ strategy users also switches to $D$ strategy. Due to this whole process fraction of $R$ strategy users decreases, fraction of $C$ strategy users increases and fraction of $D$ strategy users remain same in the new Nash equilibrium as shown in Figure \ref{betaVF2G}.
\subsection{Reputation system with round based initial payment distributed to reputation calculator $R$ users}
In the last subsection, we have analyzed the reputation system in which initial payment is distributed among the cooperation and reputation with cooperation peers ($C$ and R strategy). Now, in this subsection, we will analyze the reputation system with round based initial payment imposed to the peers, which is distributed among only reputation (R) strategy users. The game can be defined as follows.\\
\textbf{Players}:- User1, User2\\
\textbf{Strategies}:- Reputation with cooperation (R), Cooperation (C), Defection (D)\\
\textbf{Preferences}
\begin{eqnarray}
\nonumber &&  u_i(A_i,A_{-i})= (^C l_{-i} \cdot (1- ^Rl_{-i}) + ^Cl_{i} \cdot ^Rl_{-i} \cdot ^Cl_{-i}) \cdot d \\\nonumber && - (^Cl_{i} \cdot (1- ^Rl_i) + ^Cl_{-i} \cdot ^Rl_i)\cdot a - ^Rl_i \cdot \alpha + \\ && (^Cl_i \cdot ^Rl_{-i}) \cdot \beta + ^Rl_i \cdot ^Cl_i \cdot \frac{N \cdot p}{N-n_d} -p
\end{eqnarray}
where $A_1$ and $A_2$ are the actions of player 1 and player 2 respectively. $^Cl_i$ is the cooperation level of player $i$ and $^Rl_i$ is the reputation calculation level of player $i$ respectively.\\
For $C$ (cooperation) strategy : $^Cl = 1$ and $^Rl = 0$. Because these users are always cooperating and not calculating reputation. Similarly \\
For $R$ (reputation calculation) strategy : $^Cl = 1$ and $^Rl = 1$\\
For D (defection) strategy : $^Cl = 0$ and $^Rl = 0$\\
In the preference function the first term represents the benefit of sharing, the benefit of sharing the resources can only be obtained by first user when the second user is either cooperator ($C$) or when the first player is either cooperator or reputation calculator user ($C$ and $R$) and second player is reputation calculator ($R$) user. The second term represents the cost of sharing, the cost of sharing will only be imposed when the player is either cooperator or he is reputation calculator and second player is cooperator. Third term represents the cost of reputation calculation cost which is always incurred when the first user is reputation calculator ($R$) user. Fourth term is the benefit of reputation increment. Fifth term is the benefit due to initial payment distribution. Payoff matrix of the game is shown in table \ref{Third_Game}.\\
\begin{table}[!t]
\renewcommand{\arraystretch}{1.3}
\caption{Initial Payment with Distribution To Reputation Users}
\label{Third_Game}
\centering
\begin{center}
\begin{tabular}{c c	c  c}
{ }&{R(B)} & {C (B)} & {D(B)}\\ \hline
{R (A)} & \shortstack{$d-a-\alpha+\beta+$\\ $ \frac{(N-n_r) \cdot p}{n_r}$, \\$d-a-\alpha + \beta +$\\$ \frac{(N-n_r) \cdot p}{n_r}$} & \shortstack{$d-a-\alpha +$\\ $\frac{(N-n_r) \cdot p}{n_r}$,\\$d-a+\beta-p$} & {$\frac{(N-n_r) \cdot p}{n_r} -\alpha$,$-p$}\\ \hline
{C (A)} & \shortstack{$d-a+\beta - p$,\\$d-a-\alpha +$\\ $\frac{(N-n_r) \cdot p}{n_r}$} & \shortstack{$d-a-p$,\\$d-a-p$} & {$-a-p$,$d-p$}\\ \hline
{D (A)} & {$-p$,$\frac{(N-n_r) \cdot p}{n_r} -\alpha$} & {$d-p$,$-a-p$} & { $-p$,$-p$}\\ \hline
\end{tabular}
\end{center}
\end{table}
\textbf{Analysis} In this game\\
\begin{equation}
\label{G3ESScond}
\begin{split}
if\ \ \ p > \frac{n_r \cdot \alpha}{N} \\
\end{split}
\end{equation}
then we claim that reputation ($R$) strategy will be pure strategic strict Nash equilibrium and hence  evolutionarily stable.
Where $n_r$ is the number of $R$ strategy users in the population.\\
If (\ref{G3ESScond}) is not fulfilled and
\begin{equation}
\label{G3ESScond_2}
\begin{split}
if\ \ \ \ p < \frac{n_r \cdot \alpha}{N} \\
\end{split}
\end{equation}
then we claim that $D$ strategy is a pure strategy Nash equilibrium, and hence evolutionarily stable.

The argument for strategy profile ($R$,$R$) as strict Nash equilibrium, given condition (\ref{G3ESScond}), is as follows.\\
We can observe that if condition (\ref{G3ESScond}) is true, then $U_{1}(R,R) > U_{1}(C,R)$ and $U_1(R,R) \geq U_1(D,R)$ that means the payoff of R strategy when played with itself is always greater than or equal to other two strategy while played with $R$.\\
The argument for strategy profile ($D$,$D$) as strict Nash equilibrium, given condition (\ref{G3ESScond_2}), is as follows.\\
We can observe that if condition (\ref{G3ESScond_2}) is true, then $U_{1}(D,D) > U_{1}(C,D)$ and $U_1(D,D) \geq U_1(R,D)$ that means the payoff of $D$ strategy when played with itself is always greater than or equal to other two strategy while played with $D$.\\

Now we will compute the mixed strategy Nash equilibrium profile. For this the expected payoff of each strategy can be written as
\begin{subequations}
\begin{align}
\nonumber P_R &= d \cdot (x_R + x_C) - a \cdot (x_R + x_C) + \beta \cdot x_R \\ & + p \cdot \frac{N-n_r}{n_r} - \alpha \\
P_C &= d \cdot (x_R + x_C) + \beta \cdot x_R - a - p \\
P_D &= d \cdot x_C - p
\end{align}
\end{subequations}
In this game if condition (\ref{G3ESScond}) is fulfilled, then no mixed strategy equilibrium presents as $R$ is strictly dominating strategy. If this condition is not fulfilled, then we check for multiple equilibrium in the system. Let us first take the combination of $C$ and $D$ strategies. In this combination $D$ strategy always dominates the $C$ strategy, hence no mixed strategy equilibrium presents.

 If we take the combination of $R$ and $D$ strategy, then if $p = \frac{n_r \cdot \alpha}{N}$, then ($D$,$D$) will be pure strategic weak Nash equilibrium and ($R$,$R$) will be pure strategic strict Nash equilibrium. In this condition no mixed strategy equilibrium presents. If the condition is $p < \frac{n_r \cdot \alpha}{N}$, then the negative payoff mixed strategic Nash equilibrium presents but this is of no use to the system designers.
 
  Now we take the combination of $R$ and $C$ strategies. In this if $p = \frac{n_r \cdot \alpha}{N}$, then drift occurs among these two strategies and it will be continued till the expected payoff of $D$ strategy will be less than the expected payoff to these strategies,
\begin{subequations}
\begin{align}
\nonumber & x_R \cdot (d-a + \beta -\alpha +\frac{N-n_r}{n_r} \cdot p) + (1-x_R)\cdot(d-a -\alpha \\ &+\frac{N-n_r}{n_r} \cdot p) >
 x_R \cdot (-p)  + (1-x_R) \cdot (d-p)
\end{align}
\end{subequations} 
By using above equation we got
 \begin{equation}
x_R > \frac{a}{d+\beta}
 \end{equation}
 This means that when $R$ strategy users are more in the system, then playing $D$ strategy will not be lucrative. In the combination of $R$ and $C$  strategy if $p < \frac{n_r \cdot \alpha}{N}$,
 then $C$ strategy dominates $R$ strategy and so no mixed strategy equilibrium presents in this condition. Now let us take the combination of all three strategies for the mixed strategy equilibrium.
For this, the equality is,
\begin{eqnarray}
\nonumber && (d-a-\alpha + \beta + \frac{(N-n_r) \cdot p}{n_r}) x_{R}  + (d-a-\alpha \\ \nonumber && + \frac{(N-n_r) p}{n_r})  \cdot x_{C} + (\frac{(N-n_r) \cdot p}{n_r} - \alpha)(1-x_{R}-x_{C}) \\
\nonumber &&  = (d-a + \beta -p) x_{R}  + (d-a-p) \cdot x_{C} + \\
\nonumber && (-a-p)(1 -x_{R} -x_{C}) =(-p) \cdot x_{R} + (d-p) \cdot x_{C} 
\\ && + (1-x_{R}-x_{C}) \cdot (-p)
\end{eqnarray} 
By this equality we got 
\begin{subequations}
\label{MNE_game3}
\begin{align}
x_{R} & = \frac{a}{d+\beta} \\
x_{C} & = 1 - \frac{\alpha - p \cdot \frac{N}{n_r} }{a} - \frac{a}{d+\beta}\\
x_{D} & = \frac{\alpha - p \cdot \frac{N}{n_r} }{a}
\end{align}
\end{subequations}
In this mixed strategy equilibrium described by (\ref{MNE_game3}), following things can be observed:-
\begin{itemize}
\item With the increment in `cost of the reputation calculation' $\alpha$, $D$ strategy users increases, C strategy users decreases and $R$ strategy users remains same in resulting mixed strategy equilibrium
\item With the increment in `benefit of reputation increment' $\beta$, the fraction of $R$ and $D$ strategy users decreases whereas fraction of $C$ strategy users increases
\item With the increment in the `initial payment' $p$, the fraction of $D$ strategy users decreases, whereas the fraction of $C$ strategy users increases
\item With the increment in `cost of sharing' $a$, $C$ strategy users decreases, $R$ and $D$ strategy users increases in resulting mixed strategy equilibrium.
\end{itemize}
The reasoning behind first observation is that as `cost of reputation calculation' $\alpha$ increases, then the payoff to $R$ strategy users decreases and therefore the $R$ strategy becomes less lucrative to the users so they switches to C strategy and $D$ strategy users. As $R$ strategy users decreases and $C$ strategy users increases, then the payoff to $C$ strategy users also decreases and payoff to $D$ strategy users increases. Due to this the $C$ strategy users also switches to $D$ strategy users. As $D$ strategy users increases the payoff to $C$ strategy users decreases more and comes to lower than $R$ strategy. Due to this $C$ strategy users now switches to $R$ strategy till the payoff of all the strategy equalizes. Due to this in new equilibrium $D$ strategy users increases, $C$ strategy users decreases and $R$ strategy users remains same as shown in figure \ref{alphaVF1G}.

The reasoning behind second observation is that as the `benefit of reputation increment' $\beta$ increases the payoff to $R$ and $C$ strategy users increases whereas the payoff to $D$ strategy users remains same due to this $D$ strategy users switches to $R$ and $C$ strategy users. As the fraction of $D$ strategy users decreases and fraction of $R$ strategy users increases the benefit of payment to $R$ strategy users decreases so they also switches to $C$ strategy users. This whole process continues till the payoff to all three strategies equalizes. This results in the increment to the $C$ strategy fraction and decrement in the $R$ and $D$ strategy fraction of population.

The reasoning behind third observation is that as the `initial payment' $p$ increases the payoff to $R$ strategy users increases whereas the payoff to $C$ and $D$ strategy users decreases. Due to this the $C$ and $D$ strategy users switches to $R$ strategy users. As $R$ strategy users increases the payoff to $C$ strategy increases due to this $R$ strategy users switches to $C$ strategy users till the payoff to all three strategies equalizes. This process results in increment in the fraction of $C$ strategy, decrement in the fraction of $D$ strategy and remain same in the fraction of $R$ strategy.

The reasoning behind fourth observation is that as the `cost of sharing' $a$ increases the payoff to $C$ strategy users and $R$ strategy users decreases and payoff to $D$ strategy users remains constant. This results in switching of $C$ and $R$ strategy users to $D$ strategy. As the fraction of $C$ strategy users decreases this results in the decrement the payoff to $D$ strategy users and increment the payoff to $R$ strategy users as they are getting benefit from initial payment. So now the users switch to $R$ strategy users till the payoff to all three strategies equalizes and in new equilibrium fraction of the $D$ and $R$ strategy users increases whereas the fraction of $C$ strategy users decreases.
\begin{figure}
  \subfloat[]{%
  \begin{minipage}{0.5\linewidth}
  \label{bVF3G}
  \includegraphics[width=1\linewidth]{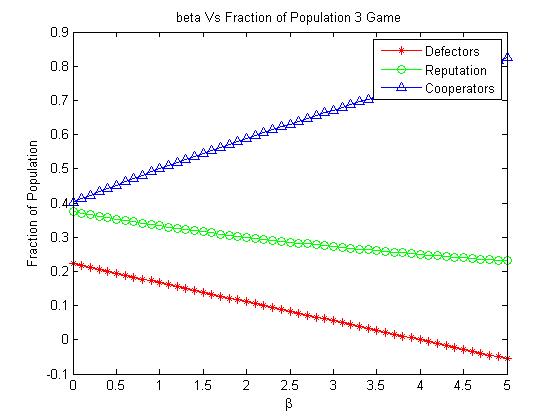}\hfill
  \end{minipage}%
  }
  \subfloat[]{%
  \begin{minipage}{0.5\linewidth}
  \label{aVsF3G}
  \includegraphics[width=1\linewidth]{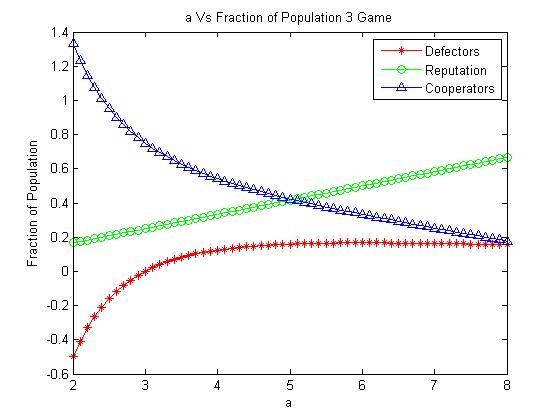}\hfill
  \end{minipage}%
  }\par
  \subfloat[]{%
  \begin{minipage}{0.5\linewidth}
  \label{pVF3G}
  \includegraphics[width=1\linewidth]{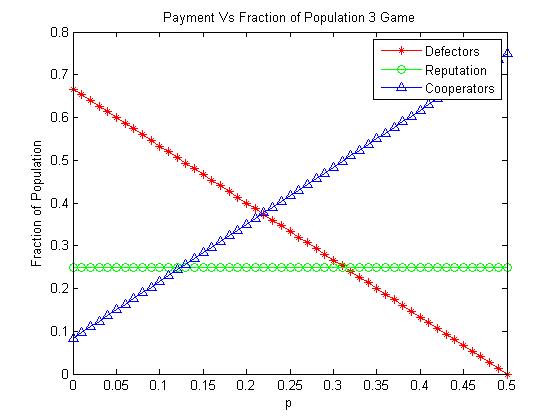}
  \end{minipage}%
  }
   \subfloat[]{%
  \begin{minipage}{0.5\linewidth}
  \label{alphaVF3G}
  \includegraphics[width=1\linewidth]{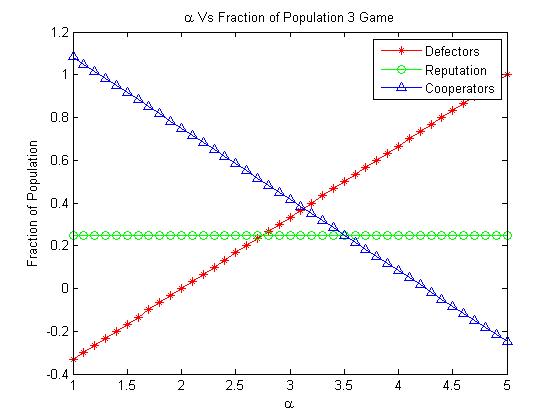}\hfill
  \end{minipage}%
  }\par
  \caption{Fraction of population in mixed Nash equilibrium for the network (a) vs benefit of reputation increment $\beta$ with other parameters \{$p=0.5$, $\alpha=2$, $a=3$, $d=8$\} (b) vs cost of sharing $a$ with other parameters \{$p=0.5$, $\alpha=2$, $d=8$, $\beta=4$\} (c) vs round based initial payment $p$ with other parameters \{$\alpha=2$, $a=3$, $d=8$, $\beta=4$\} (d) vs cost of reputation calculation $\alpha$ with other parameters \{$p=0.5$, $a=3$, $d=8$, $\beta=4$\}}
\end{figure}
\section{Numerical Analysis of different models of Reputation Game}
All the above three explained system model is analyzed by simulation as well. By simulation, we have shown the final evolution of the system. The simulation experiments have been conducted for 10000 nodes. We assume the fully connected topology of the network in which any two peer in the network can interact with each other at random in a large and well-mixed population. On the part of each user, system constitutes three strategy viz. $R$ (Reputation calculation with cooperation), $C$ (Cooperation) and $D$ (Defection). To show the relationship between the final evolved fraction and the initial fraction of different strategy users, we have taken different initial fractions of population for different strategy users and plotted their evolution separately. The evolution process contains the repetition of three phases viz. selection phase, transaction phase and reproduction phase. Initially each node selects any of the other node for pairwise interaction with equal probability so the probability that user will interact with any other strategy user is the fraction of that strategy users in the population. This phase is called selection phase. After this phase each node simultaneously calculates its payoff using the utility function based on game. This phase is called transaction phase. After each transaction, there is a reproduction phase in which all users imitate any other strategy with the probability proportional to the difference between the strategy's expected payoff and the population expected payoff. In our system model we assume that each node has the knowledge of all his neighbor's payoff and strategy. So the nodes adopts a new strategy according to the natural selection. For the simulation we also chooses the parameter values viz. $d$ (benefit of sharing), $a$ (cost of sharing), $\alpha$ (cost of reputation calculation), $\beta$ (benefit of reputation increment) and $p$ (initial payment). In the selection of the parameter values we follow constraints that is necessary and sufficient for modeling this game viz. the `cost of sharing' $a$ should always be less than or equal to the `benefit of the shared resources' $d$ and greater than `cost of reputation calculation' $\alpha$. We have examined these parameters for different values in ordinal fashion and observed that final evolution is still same.
\subsection{First Reputation Game}
\begin{figure}
  \subfloat[]{%
  \begin{minipage}{0.5\linewidth}
  \label{FGNrca}
  \includegraphics[width=1\linewidth]{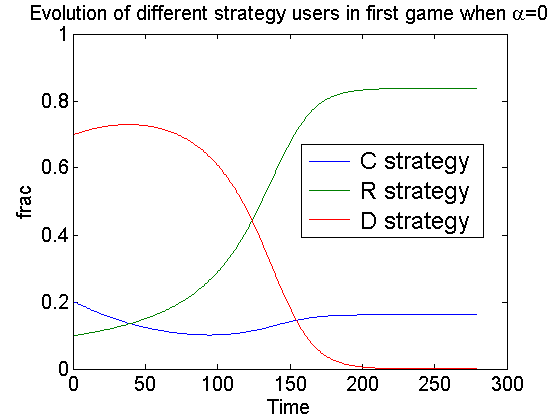}\hfill
  \end{minipage}%
  }
  \subfloat[]{%
  \begin{minipage}{0.5\linewidth}
  \label{FGNrcb}
  \includegraphics[width=1\linewidth]{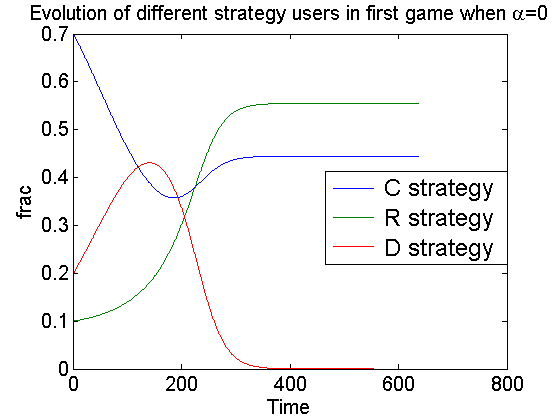}\hfill
  \end{minipage}%
  }\par
  \subfloat[]{%
  \begin{minipage}{0.5\linewidth}
  \label{FGNrcc}
  \includegraphics[width=1\linewidth]{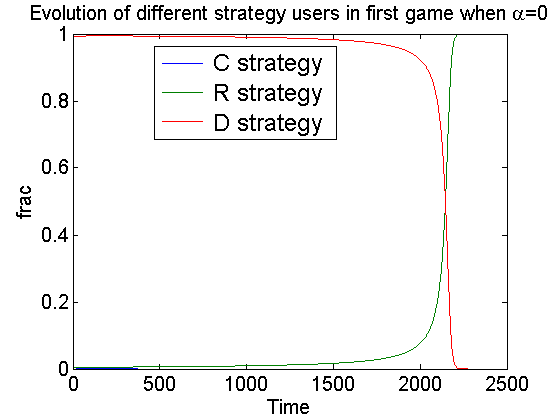}
  \end{minipage}%
  }
   \subfloat[]{%
  \begin{minipage}{0.5\linewidth}
  \label{FGNrcd}
  \includegraphics[width=1\linewidth]{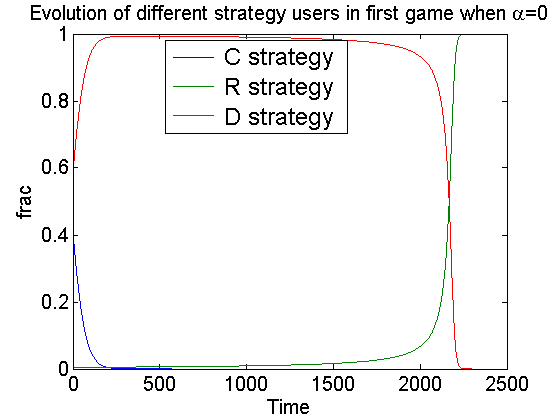}\hfill
  \end{minipage}%
  }\par
  \subfloat[]{%
  \begin{minipage}{0.5\linewidth}
  \label{FGNrce}
  \includegraphics[width=1\linewidth]{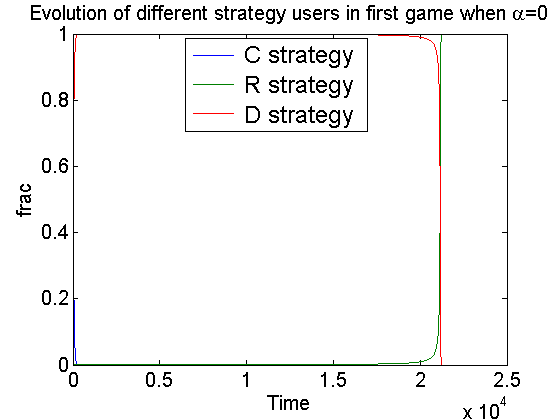}\hfill
  \end{minipage}%
  }
  \subfloat[]{%
  \begin{minipage}{0.5\linewidth}
  \label{FGNrcf}
  \includegraphics[width=1\linewidth]{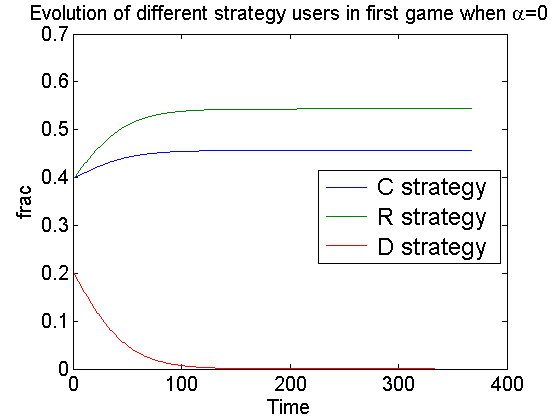}\hfill
  \end{minipage}%
  }\par
  \caption{Round vs Fraction of population for the network with d=8, a=3, $\beta=4$, $p=0.5$, $\alpha=0$}
\end{figure}
\subsubsection{Reputation cost $\alpha$ negligible}
In first scenario we have taken $\alpha$ negligible. We run our simulation for different initial fractions of $R$, $C$ and $D$ strategies and fixed parameter values ($d=8$, $a=3$, $\beta=4$, $\alpha=0$). Figure \ref{FGNrca} and \ref{FGNrcb} shows that the final fraction of $R$ and $C$ strategy depends on the initial fraction of the strategies, which substantiate the theoretical analysis of the first game. We observe that as $\frac{\alpha}{a}$ is zero so the payoff to $R$ strategy users is greater than the payoff to $C$ strategy until $D$ strategy user's fraction are greater than zero and becomes equal to the payoff of $C$ strategy users when all $D$ strategy users dies out. At first we run the simulation with 0.1, 0.2, 0.7 fraction of $R$, $C$ and $D$ strategies users respectively (figure \ref{FGNrca}). As mentioned earlier, in the selection phase user selects other user with equal probability so the probability that it will meet with $R$ strategy users is 0.1, with $C$ strategy users is 0.2 and with $D$ strategy users is 0.7. Then in transaction phase each node simultaneously calculates the payoff. Node selects other strategy with probability proportional to the difference between his neighbor's payoff and his payoff. Each node imitates to the higher payoff strategy with positive probability. In this scenario initially as $x_R<\frac{a}{d+\beta}$ i.e., $0.1<0.25$ and $x_R>\frac{a(x_R+x_C)+\alpha}{d+\beta}$ i.e., $0.1>0.075$ , results in the expected payoff order as $P_R > P_D > P_C$, so $C$ strategy imitates to $D$ and $R$ strategy whereas $D$ strategy imitates to $R$ strategy. This results in increment in the $R$ strategy and $D$ strategy fraction and decrement in $C$ strategy fraction initially. As the fraction of $R$ strategy users increases and becomes greater than $0.25$, then the expected payoff of $D$ strategy users comes to lower than $C$ strategy users which results in payoff order as $P_R > P_C > P_D$ and so from now $D$ strategy users imitates to $R$ and $C$ strategy, and $C$ strategy users imitates to only $R$ strategy. This results in increment to $R$ and $C$ strategy users whereas decrement in $D$ strategy fraction. After this as $D$ strategy fraction becomes zero which is equal to $\frac{\alpha}{a}$, then the payoff order becomes $P_R=P_C>P_D$ which further results in constant fraction of all three strategies. We observe that in this scenario as discussed in the model 1, in final evolution system composed with only $R$ and $C$ strategy users as in figure \ref{FGNrca}. After this we run the simulation with 0.1, 0.7, 0.2 initial fraction of $R$, $C$ and $D$ strategies users respectively (figure \ref{FGNrcb}). The same process as of explained earlier again happens but in this scenario the fraction of $C$ strategy users are almost greater than $0.35$ when the fraction of $R$ strategy becomes greater than $\frac{a}{d+\beta}$. So in final evolution again there is only $R$ and $C$ strategy but in this $C$ strategy users are higher than previous. In third initial setting, we took very less $R$ and $C$ user fraction i.e., $0.002$ and $0.001$ respectively and we observe that $C$ strategy dies out before the fraction of $R$ strategy becomes greater than $\frac{a}{d+\beta}$ so only $R$ strategy users remains in the final evolution. With this simulation scenario, it can be observed that as $\frac{a(x_R+x_C)+\alpha}{d+\beta}$ is zero when all the population imitates to $D$ strategy, therefore the payoff of $R$ and $D$ strategy is equal when $x_R,x_C=0$ and $x_D=1$. As $x_R$ slightly increases, the payoff of $R$ strategy users becomes greater than $D$ strategy users and $D$ strategy users imitates to $R$ strategy as in figure \ref{FGNrcc} and \ref{FGNrcd} that even with very small initial fraction from mutation, $R$ strategy is there in the final evolution. In figure \ref{FGNrcc}, \ref{FGNrcd} and \ref{FGNrce} with initial $R$ strategy fraction as 0.002, 0.005 and 0.0005 respectively, it can also be observed that as the initial fraction of $R$ strategy decreases, the final evolution time of the system increases. In figure \ref{FGNrcf} from beginning, the fraction of $D$ strategy users decrease because from beginning initial fraction of $R$ strategy remains greater than $0.25$ value.
\begin{figure}
  \subfloat[]{%
  \begin{minipage}{0.5\linewidth}
  \label{FGNa}
  \includegraphics[width=1\linewidth]{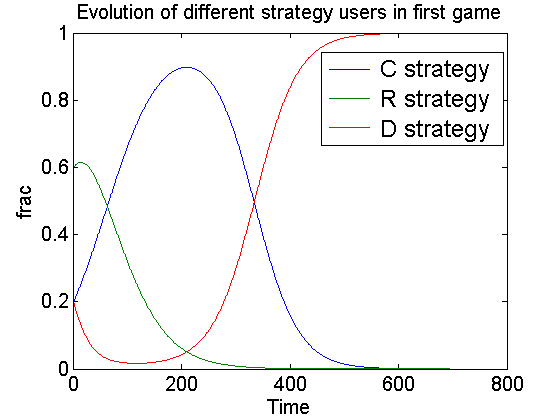}\hfill
  \end{minipage}%
  }
  \subfloat[]{%
  \begin{minipage}{0.5\linewidth}
  \label{FGNb}
  \includegraphics[width=1\linewidth]{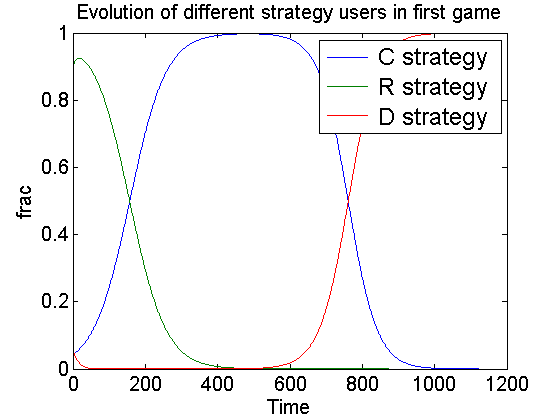}\hfill
  \end{minipage}%
  }\par
  \subfloat[]{%
  \begin{minipage}{0.5\linewidth}
  \label{FGNc}
  \includegraphics[width=1\linewidth]{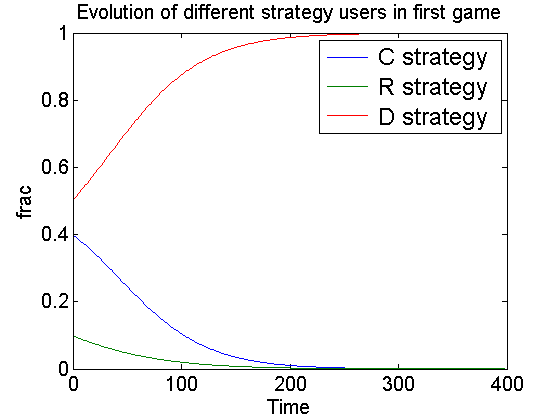}
  \end{minipage}%
  }
   \subfloat[]{%
  \begin{minipage}{0.5\linewidth}
  \label{FGNd}
  \includegraphics[width=1\linewidth]{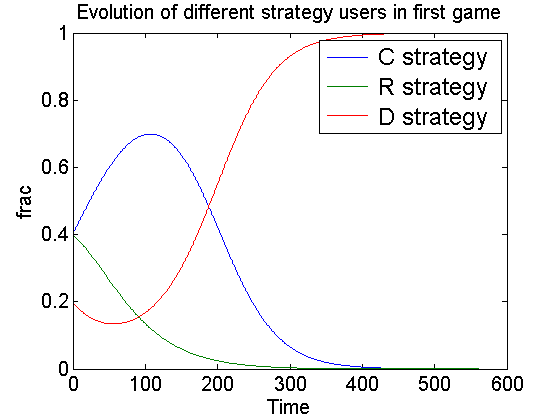}\hfill
  \end{minipage}%
  }\par
  \caption{Round vs Fraction of population for the network with d=8, a=3, $\beta=4$, $\alpha=2$}
  \label{EvoluionGame1}
\end{figure}
\subsubsection{Reputation cost $\alpha$ is not negligible}
 In the second scenario $\alpha$ is not negligible. Results of this scenario also substantiate the theoretical results of Game 1. First we have taken the fraction of different strategies as 0.6, 0.2, 0.2 fraction of $R$, $C$ and $D$ strategies users respectively as shown in figure \ref{FGNa}. Initially when the $R$ strategy users are more than $\frac{a}{d+\beta}$ and $\frac{a(x_R+x_C) + \alpha}{d+\beta}$, and $D$ strategy users are lesser than $\frac{\alpha}{a}$ so the order of expected payoff becomes $P_C>P_R>P_D$. Due to this initially $R$ strategy users imitates to $C$ strategy users whereas $D$ strategy users imitates to $R$ and $C$ strategy users. The fraction of $R$ strategy users increases till the expected payoff of $R$ strategy remains greater than the expected payoff of population. This is the point where $x_R > \frac{\alpha}{\alpha+x_D(d+\beta-a)}$ i.e., $0.61$. After this point as the fraction of $R$ strategy decreases and comes to lower than $\frac{a}{d+\beta}$ i.e., $0.25$, then the expected payoff order becomes $P_D>P_C>P_R$ so now $R$ strategy users start to imitate to both $C$ and $D$ strategy users as shown in figure \ref{FGNa}. At this point as $C$ strategy users are more than 70\% and $D$ strategy users are lesser than 10\% so $R$ strategy users interacts with more $C$ strategy users and so they imitates to more $C$ strategy users. This continues till the expected payoff of $C$ strategy users is greater than the expected payoff of the population which is the point where $x_R > \frac{a}{(d+\beta-a)+\frac{\alpha}{x_D}}$. After this point more $R$ and $C$ strategy imitates to $D$ strategy and so $C$ strategy users also decrease and in final evolution all population imitates to the $D$ strategy. After this we run the simulation with 0.9, 0.05, 0.05 initial fraction of $R$, C and $D$ strategies users respectively (figure \ref{FGNb}). In this simulation $D$ strategy users almost dies out before the fraction of $R$ strategy users comes to lower than $\frac{a}{d+\beta}$ and only $C$ strategy users remained in the system. As system consist of most of $C$ strategy users, the payoff of $D$ strategy again start to increase but for some time until some user mutate to $D$ strategy only $C$ strategy remains in the system. As mutation takes place and some users mutates to $D$ strategy, $C$ strategy users also start to imitate $D$ strategy and finally $D$ strategy invades whole the population. After this we also run the simulation with two more different initial fraction and we found the same evolution in the system as in figure \ref{FGNc} and \ref{FGNd}.
 \begin{figure}
  \subfloat[]{%
  \begin{minipage}{0.5\linewidth}
  \label{FGNpca}
  \includegraphics[width=1\linewidth]{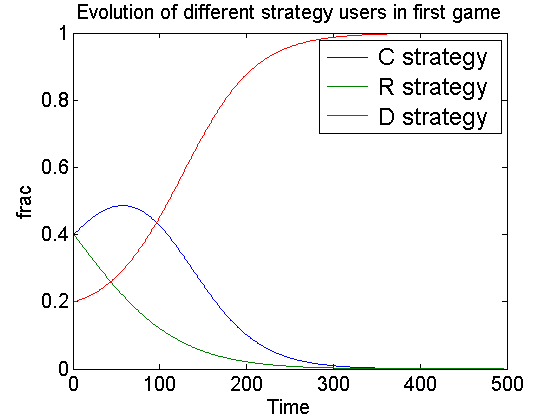}
  \end{minipage}%
  }
  \subfloat[]{%
  \begin{minipage}{0.5\linewidth}
  \label{FGNpcb}
  \includegraphics[width=1\linewidth]{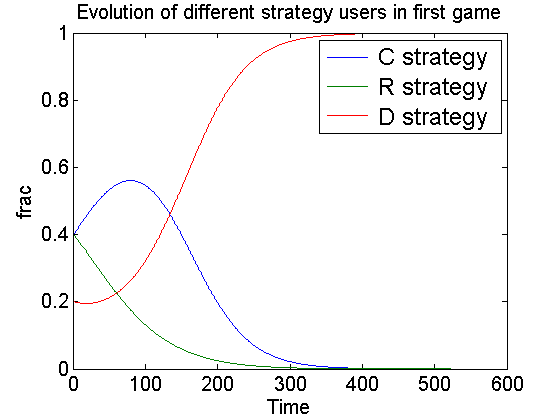}\hfill
  \end{minipage}%
  }\par
  \subfloat[]{%
  \begin{minipage}{0.5\linewidth}
  \label{FGNpcc}
  \includegraphics[width=1\linewidth]{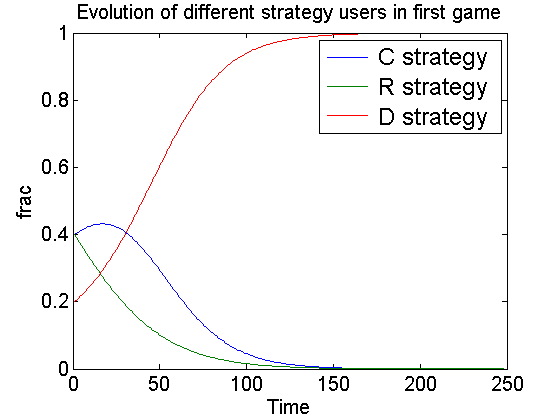}
  \end{minipage}%
  }
   \subfloat[]{%
  \begin{minipage}{0.5\linewidth}
  \label{FGNpcd}
  \includegraphics[width=1\linewidth]{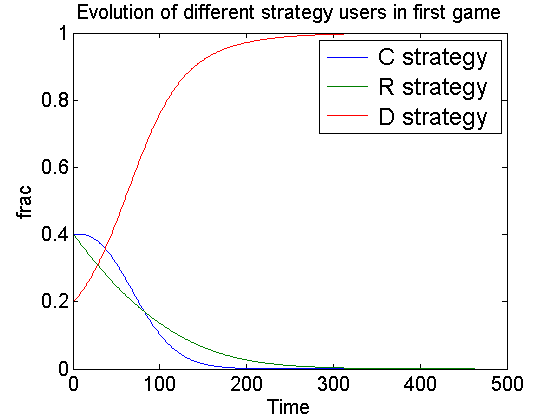}\hfill
  \end{minipage}%
  }\par
  \caption{Round vs Fraction of population for the network with parameter values (a) $d=4$, $a=2.5$, $\beta=3$, $\alpha=2$ (b) $d=6$, $a=2.5$, $\beta=3$, $\alpha=2$ (c) $d=6$, $a=5$, $\beta=3$, $\alpha=4$ (d) $d=8$, $a=5$, $\beta=1$, $\alpha=2$}
\end{figure}
 We also run the simulation with different parameter for the same initial fraction (0.4, 0.4 and 0.2 of $R$, $C$ and $D$ respectively). At first simulation we took $d=4$, $a=2.5$, $\beta=3$ and $\alpha=2$. In this setting initially as the $R$ strategy users are more than $\frac{a}{d+\beta}$ i.e., $0.35$ and lesser than $\frac{a(x_R+x_C) + \alpha}{d+\beta}$, and $D$ strategy users are lesser than $\frac{\alpha}{a}$, so expected payoff order becomes $P_C>P_D>P_R$. Due to this $R$ strategy users imitates to $C$ and $D$ strategy users. As $R$ strategy users decreases and comes to lower than $0.35$, the payoff to $D$ strategy users becomes greater than $C$ strategy users and so $C$ strategy users also start to imitate to $D$ strategy users due to this $C$ strategy users start to decrease. In this way the final evolution reaches with all $D$ strategy users in the population. In second simulation we increases the value of parameter $d$ to 6. We found that now in this setting initially more $R$ strategy users imitates to $C$ strategy, as now the $\frac{a}{d+\beta}$ is $0.27$ which is more lesser than $x_R$ than previous and so the rate of imitation is more than previous, but again in final evolution all users imitates to $D$ strategy. In third simulation we increases both cost of sharing $a$ and cost of reputation calculation $\alpha$ to 5 and 4 respectively. Due to this the increment in $C$ strategy population stops early from previous and users start to imitate to $D$ strategy earlier than previous setting. In fourth setting as we decreases the benefit of reputation increment the $C$ strategy users start to decrease earlier than previous setting. We found that even with different parameter setting the final evolution is same.
\subsection{Numerical Analysis of Second Reputation Game}
\begin{figure}
  \subfloat[]{%
  \begin{minipage}{0.5\linewidth}
  \label{SGNa}
  \includegraphics[width=1\linewidth]{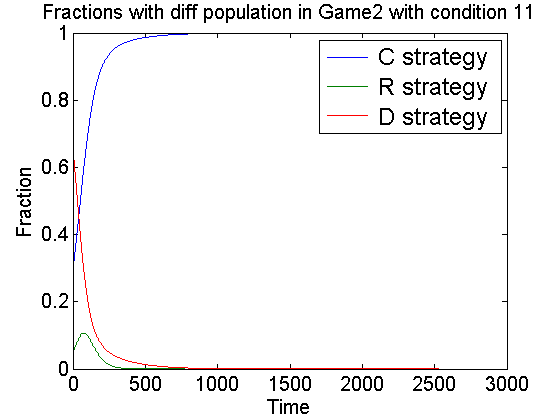}\hfill
  \end{minipage}%
  }
  \subfloat[]{%
  \begin{minipage}{0.5\linewidth}
  \label{SGNb}
  \includegraphics[width=1\linewidth]{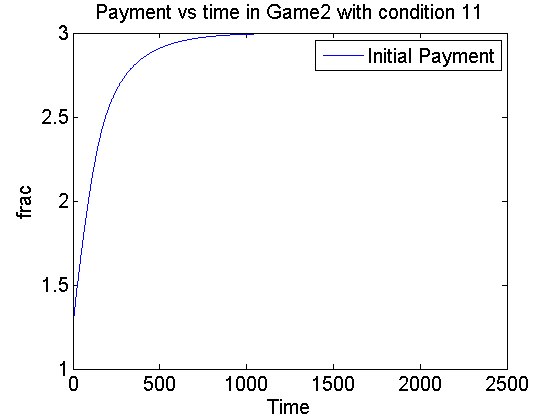}\hfill
  \end{minipage}%
  }\par
  \subfloat[]{%
  \begin{minipage}{0.5\linewidth}
  \label{SGNc}
  \includegraphics[width=1\linewidth]{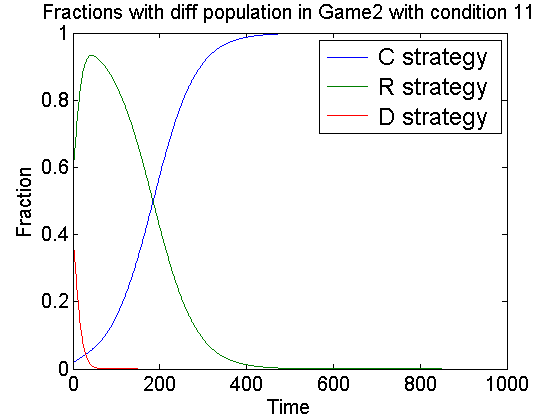}
  \end{minipage}%
  }
   \subfloat[]{%
  \begin{minipage}{0.5\linewidth}
  \label{SGNd}
  \includegraphics[width=1\linewidth]{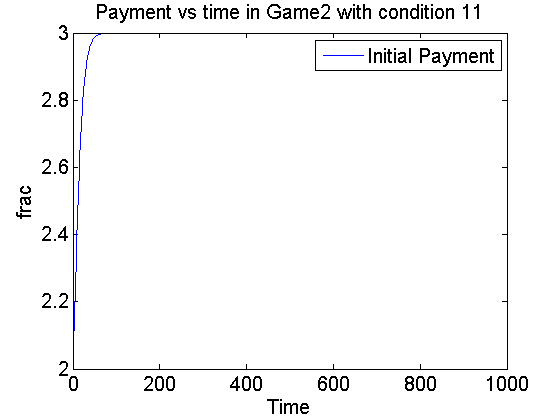}\hfill
  \end{minipage}%
  }\par
  \subfloat[]{%
  \begin{minipage}{0.5\linewidth}
  \label{SGNe}
  \includegraphics[width=1\linewidth]{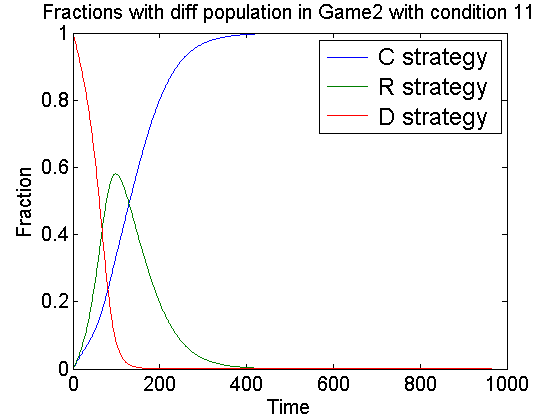}\hfill
  \end{minipage}%
  }
  \subfloat[]{%
  \begin{minipage}{0.5\linewidth}
  \label{SGNf}
  \includegraphics[width=1\linewidth]{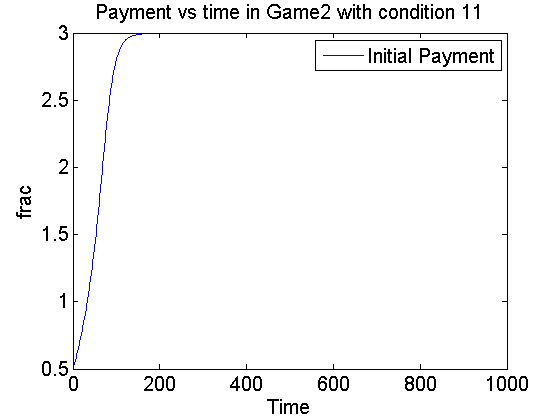}\hfill
  \end{minipage}%
  }\par
  \caption{Round vs Fraction of population with corresponding varying payment vs time for the network with $d=8$, $a=3$, $\beta=4$, $\alpha=2$}
\end{figure}
In first simulation scenario, round based initial payment $p$ varies according to equation \ref{G2ESScond} for different initial fractions of strategies. We observe that when $D$ strategy users are more, then less $p$ is required and as $D$ strategy users decreases, $p$ increases, this is due to the fact that as $D$ strategy users are more, then $p$ is distributed among less users. Although initial payment in the defection free population is high but this initial payment is given back to the users during redistribution. The initial payment is not distributed back only to the users who defects and their part is distributed among the users who cooperates in the form of reward. First we run the simulation with fraction 0.05, 0.3 and 0.65 of $R$, $C$ and $D$ strategy as shown in figure \ref{SGNa}. As in this game if $x_D>\frac{\alpha}{a}$ then $P_R > P_C$ and initially the fraction of $D$ strategy users are almost equal to $\frac{\alpha}{a}$ i.e., $0.66$, therefore the payoff to $C$ strategy users are almost equal to $R$ strategy users. Also as if $x_R > \frac{a(1-x_D) + \alpha -\frac{p}{1-x_D}}{d+\beta}$ then $P_R > P_D$ and initially $x_R$ is greater than $\frac{a(1-x_D) + \alpha -\frac{p}{1-x_D}}{d+\beta}$ i.e $0.05 > 0.02$. Due to this the order of expected payoff becomes $P_C>P_R>P_D$ and $D$ strategy users imitates to $R$ and $C$ strategy. Therefore initially $R$ and $C$ both strategy users increase but as $D$ strategy users fraction decreases and $p$ increases then the fraction $\frac{a(1-x_D) + \alpha -\frac{p}{1-x_D}}{d+\beta}$ increases and at point when $x_D$ becomes $0.25$ than $x_R$ comes to lower than this fraction and so from now the expected payoff of $D$ strategy users becomes greater than $R$ strategy users and so $R$ strategy users also start to imitate to $D$ and $C$ strategy users. This process continues and in final evolution all users imitates to $C$ strategy. The same process repeats in other two evolution with initial fraction 0.6, 0.05 and 0.35 of $R$, $C$ and $D$ strategy respectively as shown in figure \ref{SGNb}, and with initial fraction 0.025, 0.025 and 0.95 of $R$, $C$ and $D$ strategy respectively as shown in figure \ref{SGNc}.
 In second simulation scenario the round based initial payment fulfills the condition \ref{G2ND}. This evolution is same as first game evolution figure \ref{EvoluionGame1}.
\subsection{Numerical Analysis of Third Reputation Game}
\begin{figure}
  \subfloat[]{%
  \begin{minipage}{0.5\linewidth}
  \label{TGNa}
  \includegraphics[width=1\linewidth]{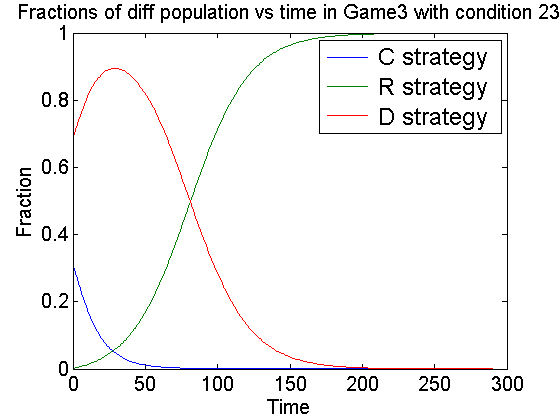}\hfill
  \end{minipage}%
  }
  \subfloat[]{%
  \begin{minipage}{0.5\linewidth}
  \label{TGNap}
  \includegraphics[width=1\linewidth]{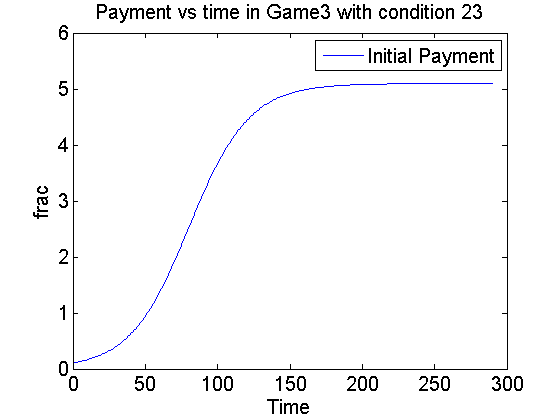}\hfill
  \end{minipage}%
  }\par
  \subfloat[]{%
    \begin{minipage}{0.5\linewidth}
    \label{TGNb}
    \includegraphics[width=1\linewidth]{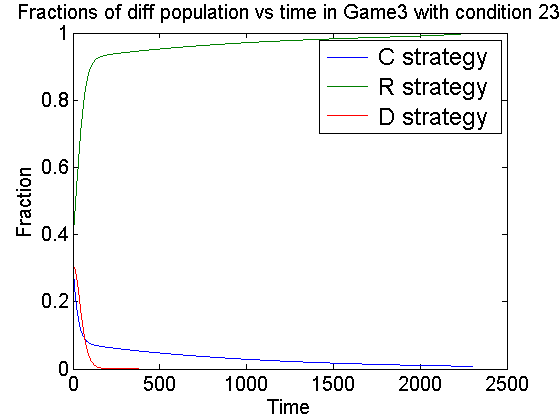}\hfill
    \end{minipage}%
    }
    \subfloat[]{%
    \begin{minipage}{0.5\linewidth}
    \label{TGNbp}
    \includegraphics[width=1\linewidth]{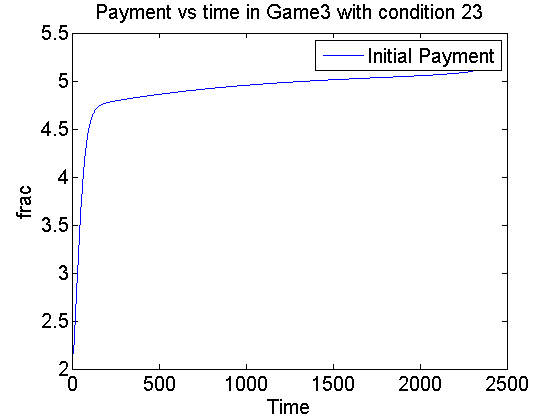}\hfill
    \end{minipage}%
    }\par
  \caption{Round vs Fraction of population with corresponding varying payment vs time for the network with $d=8$, $a=2.5$, $\beta=4$, $\alpha=2$}
\end{figure}
In this simulation scenario round based initial payment is varied according to equation \ref{G3ESScond}.   First we run the simulation for different initial fraction. At first we took fraction 0.002, 0.3 and 0.698 of $R$, $C$ and $D$ strategy respectively. As initially $R$ strategy users are lesser than $\frac{a}{d+\beta}$ so the payoff order becomes $P_R>P_D>P_C$. But as $D$ strategy are more so $C$ strategy users interact more $D$ strategy users and so they imitates to more $D$ strategy users initially. But as $R$ strategy users population increases, then $D$ strategy payoff start to decrease as $p$ is distributed among only $R$ strategic users so they got higher payoff than $C$ and $D$ strategy users and so $C$ and $D$ strategy users imitates to $R$ strategy users. In this scenario $p$ varies according to the fraction of $R$ strategy users so the payoff also varies accordingly. This results in getting more payoff to $R$ strategic users as they are getting higher payoff from $C$ strategy in the form of initial payment distribution. This process continues and in final evolution all population converges to all the $R$ strategy users as shown in figure \ref{TGNap}. We run the second simulation with 0.4, 0.3 and 0.3 fraction of $R$, $C$ and $D$ strategy as shown in figure \ref{TGNb}. In this fraction as $R$ strategy users are greater than $\frac{a}{d+\beta}$ so $D$ strategy users does not get better payoff and so both $D$ and $C$ strategy users imitates to $R$ strategy. This process continues and in final evolution again all population converges to all $R$ strategy users as in figure \ref{TGNb}.
\begin{figure}
  \subfloat[]{%
  \begin{minipage}{0.5\linewidth}
  \label{TGNpca}
  \includegraphics[width=1\linewidth]{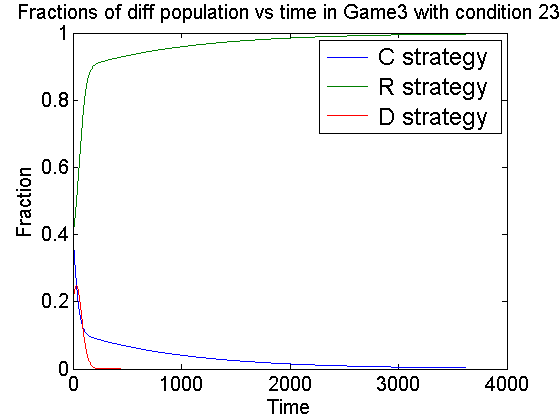}
  \end{minipage}%
  }
  \subfloat[]{%
  \begin{minipage}{0.5\linewidth}
  \label{TGNpcb}
  \includegraphics[width=1\linewidth]{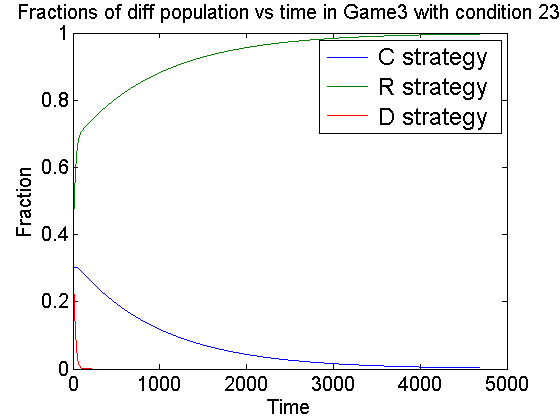}\hfill
  \end{minipage}%
  }\par
  \subfloat[]{%
  \begin{minipage}{0.5\linewidth}
  \label{TGNpcc}
  \includegraphics[width=1\linewidth]{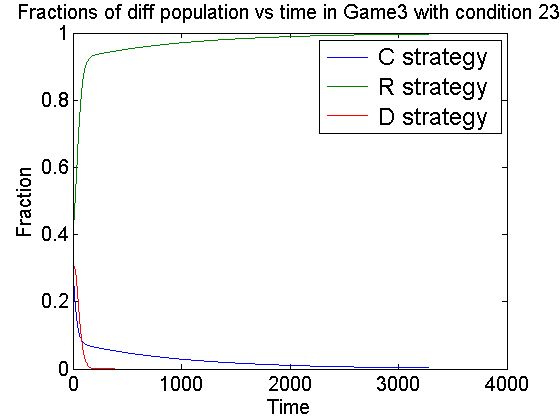}
  \end{minipage}%
  }
   \subfloat[]{%
  \begin{minipage}{0.5\linewidth}
  \label{TGNpcd}
  \includegraphics[width=1\linewidth]{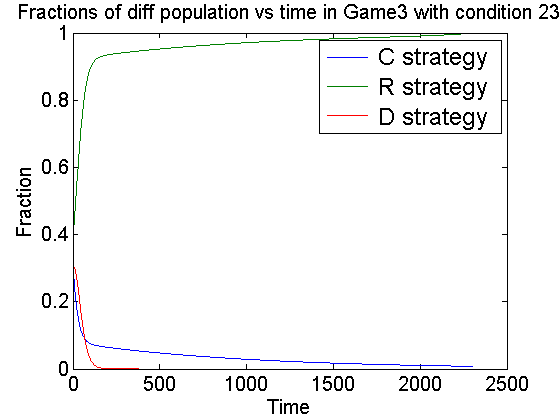}\hfill
  \end{minipage}%
  }\par
  \caption{Round vs Fraction of population for the network with parameter values (a) $d=8$, $a=7.5$, $\beta=4$, $\alpha=1$ (b) $d=8$, $a=3$, $\beta=4$, $\alpha=2.5$ (c) $d=8$, $a=7.5$, $\beta=4$, $\alpha=3$ (d) $d=8$, $a=7.5$, $\beta=4$, $\alpha=5$}
 \label{EvolutionG3PC}
\end{figure}

 Now we fix the initial fraction to 0.4,0.3 and 0.3 for $R$, $C$ and $D$ respectively and varies the parameter values as in figure \ref{EvolutionG3PC}. We observe that as the difference between $d$ and $a$ increases, $D$ strategy reduction rate increases and $C$ strategy reduction rate decrease as shown in figure \ref{TGNpca}, \ref{TGNpcb}. This is because, with the increment of this difference $C$ strategy users payoff increases. We observed that when the number of $R$ strategy users are less in the population then less $p$ is required for motivating the players to imitate $R$ strategy. But as number of $R$ strategy users becomes more, more $p$ required which is redistributed among these $R$ strategy users. Therefore the contribution of this varying round based initial payment is that, when the system consists of only $R$ strategy users, then the value of $p$ will be $\alpha$ which is distributed back to all the $R$ strategy users and so no burden of the initial payment because whatever they are paying, they are getting the same. But as any player defects then he will not get his part and also his part is distributed among other $R$ strategy users. So this mechanism is punishing users who are not calculating reputation as well as rewarding the $R$ strategy users.
\section{Discussion and Future Work}
In our analysis we have observed that if varying round based initial payment is distributed among $R$ strategy users then $R$ strategy is evolutionary stable. In this analysis $R$ strategy users fully cooperates with cooperating users and fully defects with defectors. But as the reputation of users may be in analog form and not in binary form \cite{satsiou2010reputation}, the reputation strategy should be modified accordingly. With this setting of the game, we would have the continuum of the pure strategies with all the varying level. This setting of the game would allow us to study more practical reputation system based peer-to-peer network. Furthermore the payment distribution mechanism and recognition of reputation calculator users would also needs to be further investigated. In future a reputation system may be built that will overcome these limitations.
\section{Conclusion}
We analyse the reputation game in peer-to-peer network and found that without any additional incentive, reputation strategy is not an evolutionary stable strategy. In systems, where reputation strategy is used for promoting the cooperation, even on these systems, reputation is not an evolutionary stable strategy. For making the reputation strategy as evolutionary stable strategy first varying round based initial payment has to be incorporated and then this initial payment should be distributed among $R$ strategy users. We also analysed a game in which varying initial payment is to be distributed among $C$ and $R$ strategy users and in that game cooperation strategy would be an evolutionary stable strategy for varying initial payment. We also found that whether a number of different strategies has been found for stopping free-riding using reputation, but reputation alone is not an evolutionary stable strategy.
\bibliography{reputation}

% Generated by IEEEtran.bst, version: 1.13 (2008/09/30)
\begin{thebibliography}{10}
\providecommand{\url}[1]{#1}
\csname url@samestyle\endcsname
\providecommand{\newblock}{\relax}
\providecommand{\bibinfo}[2]{#2}
\providecommand{\BIBentrySTDinterwordspacing}{\spaceskip=0pt\relax}
\providecommand{\BIBentryALTinterwordstretchfactor}{4}
\providecommand{\BIBentryALTinterwordspacing}{\spaceskip=\fontdimen2\font plus
\BIBentryALTinterwordstretchfactor\fontdimen3\font minus
  \fontdimen4\font\relax}
\providecommand{\BIBforeignlanguage}[2]{{%
\expandafter\ifx\csname l@#1\endcsname\relax
\typeout{** WARNING: IEEEtran.bst: No hyphenation pattern has been}%
\typeout{** loaded for the language `#1'. Using the pattern for}%
\typeout{** the default language instead.}%
\else
\language=\csname l@#1\endcsname
\fi
#2}}
\providecommand{\BIBdecl}{\relax}
\BIBdecl

\bibitem{adar2000free}
E.~Adar and B.~A. Huberman, ``Free riding on gnutella,'' \emph{First monday},
  vol.~5, no.~10, 2000.

\bibitem{hughes2005free}
D.~Hughes, G.~Coulson, and J.~Walkerdine, ``Free riding on gnutella revisited:
  the bell tolls?'' \emph{IEEE distributed systems online}, vol.~6, no.~6,
  2005.

\bibitem{karakaya2009free}
M.~Karakaya, I.~Korpeoglu, and {\"O}.~Ulusoy, ``Free riding in peer-to-peer
  networks,'' \emph{IEEE Internet computing}, vol.~13, no.~2, pp. 92--98, 2009.

\bibitem{karakaya2008counteracting}
M.~Karakaya, {\.I}.~K{\"o}rpeo{\u{g}}lu, and {\"O}.~Ulusoy, ``Counteracting
  free riding in peer-to-peer networks,'' \emph{Computer Networks}, vol.~52,
  no.~3, pp. 675--694, 2008.

\bibitem{feldman2005overcoming}
M.~Feldman and J.~Chuang, ``Overcoming free-riding behavior in peer-to-peer
  systems,'' \emph{ACM sigecom exchanges}, vol.~5, no.~4, pp. 41--50, 2005.

\bibitem{buragohain2003game}
C.~Buragohain, D.~Agrawal, and S.~Suri, ``A game theoretic framework for
  incentives in p2p systems,'' \emph{arXiv preprint cs/0310039}, 2003.

\bibitem{lee2003cooperative}
S.~Lee, R.~Sherwood, and B.~Bhattacharjee, ``Cooperative peer groups in nice,''
  in \emph{INFOCOM 2003. Twenty-Second Annual Joint Conference of the IEEE
  Computer and Communications. IEEE Societies}, vol.~2.\hskip 1em plus 0.5em
  minus 0.4em\relax IEEE, 2003, pp. 1272--1282.

\bibitem{dutta2003design}
D.~Dutta, A.~Goel, R.~Govindan, and H.~Zhang, ``The design of a distributed
  rating scheme for peer-to-peer systems,'' in \emph{Workshop on Economics of
  Peer-to-Peer Systems}, vol. 264, 2003, pp. 214--223.

\bibitem{papaioannou2006reputation}
T.~G. Papaioannou and G.~D. Stamoulis, ``Reputation-based policies that provide
  the right incentives in peer-to-peer environments,'' \emph{Computer
  Networks}, vol.~50, no.~4, pp. 563--578, 2006.

\bibitem{andrade2004discouraging}
N.~Andrade, F.~Brasileiro, W.~Cirne, and M.~Mowbray, ``Discouraging free riding
  in a peer-to-peer cpu-sharing grid,'' in \emph{High performance Distributed
  Computing, 2004. Proceedings. 13th IEEE International Symposium on}.\hskip
  1em plus 0.5em minus 0.4em\relax IEEE, 2004, pp. 129--137.

\bibitem{marti2004limited}
S.~Marti and H.~Garcia-Molina, ``Limited reputation sharing in p2p systems,''
  in \emph{Proceedings of the 5th ACM conference on Electronic commerce}.\hskip
  1em plus 0.5em minus 0.4em\relax ACM, 2004, pp. 91--101.

\bibitem{kamvar2003eigentrust}
S.~D. Kamvar, M.~T. Schlosser, and H.~Garcia-Molina, ``The eigentrust algorithm
  for reputation management in p2p networks,'' in \emph{Proceedings of the 12th
  international conference on World Wide Web}.\hskip 1em plus 0.5em minus
  0.4em\relax ACM, 2003, pp. 640--651.

\bibitem{xiong2004peertrust}
L.~Xiong and L.~Liu, ``Peertrust: Supporting reputation-based trust for
  peer-to-peer electronic communities,'' \emph{IEEE transactions on Knowledge
  and Data Engineering}, vol.~16, no.~7, pp. 843--857, 2004.

\bibitem{zhou2007powertrust}
R.~Zhou and K.~Hwang, ``Powertrust: A robust and scalable reputation system for
  trusted peer-to-peer computing,'' \emph{IEEE Transactions on parallel and
  distributed systems}, vol.~18, no.~4, pp. 460--473, 2007.

\bibitem{zhou2008gossiptrust}
R.~Zhou, K.~Hwang, and M.~Cai, ``Gossiptrust for fast reputation aggregation in
  peer-to-peer networks,'' \emph{IEEE Transactions on Knowledge and Data
  Engineering}, vol.~20, no.~9, pp. 1282--1295, 2008.

\bibitem{feldman2004robust}
M.~Feldman, K.~Lai, I.~Stoica, and J.~Chuang, ``Robust incentive techniques for
  peer-to-peer networks,'' in \emph{Proceedings of the 5th ACM conference on
  Electronic commerce}.\hskip 1em plus 0.5em minus 0.4em\relax ACM, 2004, pp.
  102--111.

\bibitem{kung2003differentiated}
H.~Kung and C.-H. Wu, ``Differentiated admission for peer-to-peer systems:
  incentivizing peers to contribute their resources,'' in \emph{1st Workshop on
  Economics of Peer-to-Peer Systems}, 2003.

\bibitem{meo2005rational}
M.~Meo and F.~Milan, ``A rational model for service rate allocation in
  peer-to-peer networks,'' in \emph{Proceedings 24th Annual Joint Conference of
  the IEEE Computer and Communications Societies.}, 2005.

\bibitem{satsiou2010reputation}
A.~Satsiou and L.~Tassiulas, ``Reputation-based resource allocation in p2p
  systems of rational users,'' \emph{IEEE Transactions on Parallel and
  Distributed Systems}, vol.~21, no.~4, pp. 466--479, 2010.

\bibitem{gupta2015reputation}
R.~Gupta, N.~Singha, and Y.~N. Singh, ``Reputation based probabilistic resource
  allocation for avoiding free riding and formation of common interest groups
  in unstructured p2p networks,'' \emph{Peer-to-Peer Networking and
  Applications}, pp. 1--13, 2015.

\bibitem{gupta2013avoiding}
R.~Gupta and Y.~N. Singh, ``Avoiding whitewashing in unstructured peer-to-peer
  resource sharing network,'' \emph{arXiv preprint arXiv:1307.5057}, 2013.

\bibitem{lai2003incentives}
K.~Lai, M.~Feldman, I.~Stoica, and J.~Chuang, ``Incentives for cooperation in
  peer-to-peer networks,'' in \emph{Workshop on economics of peer-to-peer
  systems}, 2003, pp. 1243--1248.

\bibitem{ma2006incentive}
R.~T. Ma, S.~Lee, J.~Lui, and D.~K. Yau, ``Incentive and service
  differentiation in p2p networks: a game theoretic approach,'' \emph{IEEE/ACM
  Transactions on Networking (TON)}, vol.~14, no.~5, pp. 978--991, 2006.

\bibitem{ma2006demand}
H.~Ma and H.-f. Leung, ``A demand and contribution based bandwidth allocation
  mechanism in p2p networks: a game-theoretic analysis,'' in \emph{20th
  International Conference on Advanced Information Networking and
  Applications-Volume 1 (AINA'06)}, vol.~1.\hskip 1em plus 0.5em minus
  0.4em\relax IEEE, 2006, pp. 1005--1010.

\bibitem{yan2007ranking}
Y.~Yan, A.~El-Atawy, and E.~Al-Shaer, ``Ranking-based optimal resource
  allocation in peer-to-peer networks,'' in \emph{IEEE INFOCOM 2007-26th IEEE
  International Conference on Computer Communications}.\hskip 1em plus 0.5em
  minus 0.4em\relax IEEE, 2007, pp. 1100--1108.

\bibitem{wang2015vpef}
Y.~Wang, A.~V. Vasilakos, and J.~Ma, ``Vpef: A simple and effective incentive
  mechanism in community-based autonomous networks,'' \emph{IEEE Transactions
  on Network and Service Management}, vol.~12, no.~1, pp. 75--86, 2015.

\bibitem{seredynski2009evolutionary}
M.~Seredynski and P.~Bouvry, ``Evolutionary game theoretical analysis of
  reputation-based packet forwarding in civilian mobile ad hoc networks,'' in
  \emph{Parallel \& Distributed Processing, 2009. IPDPS 2009. IEEE
  International Symposium on}.\hskip 1em plus 0.5em minus 0.4em\relax IEEE,
  2009, pp. 1--8.

\bibitem{wang2011effectiveness}
Y.~Wang, A.~Nakao, A.~V. Vasilakos, and J.~Ma, ``On the effectiveness of
  service differentiation based resource-provision incentive mechanisms in
  dynamic and autonomous p2p networks,'' \emph{Computer Networks}, vol.~55,
  no.~17, pp. 3811--3831, 2011.

\end{thebibliography}
\bibliographystyle{IEEEtran}

\end{document}